\documentclass[twocolumn]{aastex62}

\graphicspath{{./}{figures/}}

\def\ergs{${\rm erg\,s^{-1}}$}

\usepackage{color,subfigure}

\newcommand{\javelin}{{\tt JAVELIN}\,}

\usepackage{ulem}
\usepackage{amsmath,bm}

\shorttitle{Dust Reverberation Mapping}
\shortauthors{Yang et al.}

\begin{document}

\title{Dust Reverberation Mapping in Distant Quasars from Optical and Mid-Infrared Imaging Surveys}

\correspondingauthor{Qian Yang, Yue Shen}
\email{qiany@illinois.edu, shenyue@illinois.edu}

\author[0000-0002-6893-3742]{Qian Yang}
\affiliation{Department of Astronomy, University of Illinois at Urbana-Champaign, Urbana, IL 61801, USA}

\author[0000-0003-1659-7035]{Yue Shen}
\altaffiliation{Alfred P. Sloan Research Fellow}
\affiliation{Department of Astronomy, University of Illinois at Urbana-Champaign, Urbana, IL 61801, USA}
\affiliation{National Center for Supercomputing Applications, University of Illinois at Urbana-Champaign, Urbana, IL 61801, USA}

\author[0000-0003-0049-5210]{Xin Liu}
\affiliation{Department of Astronomy, University of Illinois at Urbana-Champaign, Urbana, IL 61801, USA}
\affiliation{National Center for Supercomputing Applications, University of Illinois at Urbana-Champaign, Urbana, IL 61801, USA}

\author{Michel Aguena}
\affiliation{Departamento de F\'isica Matem\'atica, Instituto de F\'isica, Universidade de S\~ao Paulo, CP 66318, S\~ao Paulo, SP, 05314-970, Brazil}
\affiliation{Laborat\'orio Interinstitucional de e-Astronomia - LIneA, Rua Gal. Jos\'e Cristino 77, Rio de Janeiro, RJ - 20921-400, Brazil}

\author[0000-0002-0609-3987]{James Annis}
\affiliation{Fermi National Accelerator Laboratory, P. O. Box 500, Batavia, IL 60510, USA}

\author{Santiago Avila}
\affiliation{Instituto de Fisica Teorica UAM/CSIC, Universidad Autonoma de Madrid, 28049 Madrid, Spain}

\author{Manda Banerji}
\affiliation{Institute of Astronomy, University of Cambridge, Madingley Road, Cambridge CB3 0HA, UK}
\affiliation{Kavli Institute for Cosmology, University of Cambridge, Madingley Road, Cambridge CB3 0HA, UK}

\author{Emmanuel Bertin}
\affiliation{CNRS, UMR 7095, Institut d'Astrophysique de Paris, F-75014, Paris, France}
\affiliation{Sorbonne Universit\'es, UPMC Univ Paris 06, UMR 7095, Institut d'Astrophysique de Paris, F-75014, Paris, France}

\author[0000-0002-8458-5047]{David Brooks}
\affiliation{Department of Physics \& Astronomy, University College London, Gower Street, London, WC1E 6BT, UK}

\author{David Burke}
\affiliation{Kavli Institute for Particle Astrophysics \& Cosmology, P. O. Box 2450, Stanford University, Stanford, CA 94305, USA}
\affiliation{SLAC National Accelerator Laboratory, Menlo Park, CA 94025, USA}

\author[0000-0003-3044-5150]{Aurelio Carnero Rosell}
\affiliation{Instituto de Astrof\'isica de Canarias, E-38205 La Laguna, Tenerife, Spain}
\affiliation{Departamento de Astrof\'\i{}sica, Universidad de La Laguna, E-38206 La Laguna, Tenerife, Spain}

\author[0000-0002-4802-3194]{Matias Carrasco Kind}
\affiliation{Department of Astronomy, University of Illinois at Urbana-Champaign, Urbana, IL 61801, USA}
\affiliation{National Center for Supercomputing Applications, University of Illinois at Urbana-Champaign, Urbana, IL 61801, USA}

\author{Luiz da Costa}
\affiliation{Laborat\'orio Interinstitucional de e-Astronomia - LIneA, Rua Gal. Jos\'e Cristino 77, Rio de Janeiro, RJ - 20921-400, Brazil}
\affiliation{Observat\'orio Nacional, Rua Gal. Jos\'e Cristino 77, Rio de Janeiro, RJ - 20921-400, Brazil}

\author[0000-0001-8318-6813]{Juan De Vicente}
\affiliation{Centro de Investigaciones Energ\'eticas, Medioambientales y Tecnol\'ogicas (CIEMAT), Madrid, Spain}

\author[0000-0002-0466-3288]{Shantanu Desai}
\affiliation{Department of Physics, IIT Hyderabad, Kandi, Telangana 502285, India}

\author[0000-0002-8357-7467]{H. Thomas Diehl}
\affiliation{Fermi National Accelerator Laboratory, P. O. Box 500, Batavia, IL 60510, USA}

\author{Peter Doel}
\affiliation{Department of Physics \& Astronomy, University College London, Gower Street, London, WC1E 6BT, UK}

\author[0000-0002-2367-5049]{Brenna Flaugher}
\affiliation{Fermi National Accelerator Laboratory, P. O. Box 500, Batavia, IL 60510, USA}

\author{Pablo Fosalba}
\affiliation{Institut d'Estudis Espacials de Catalunya (IEEC), 08193 Barcelona, Spain}

\author[0000-0003-4079-3263]{Josh Frieman}
\affiliation{Fermi National Accelerator Laboratory, P. O. Box 500, Batavia, IL 60510, USA}
\affiliation{Kavli Institute for Cosmological Physics, University of Chicago, Chicago, IL 60637, USA}

\author[0000-0002-9370-8360]{Juan Garcia-Bellido}
\affiliation{Instituto de Fisica Teorica UAM/CSIC, Universidad Autonoma de Madrid, 28049 Madrid, Spain}

\author[0000-0001-6942-2736]{David Gerdes}
\affiliation{Department of Astronomy, University of Michigan, Ann Arbor, MI 48109, USA}
\affiliation{Department of Physics, University of Michigan, Ann Arbor, MI 48109, USA}

\author[0000-0003-3270-7644]{Daniel Gruen}
\affiliation{Department of Physics, Stanford University, 382 Via Pueblo Mall, Stanford, CA 94305, USA}
\affiliation{Kavli Institute for Particle Astrophysics \& Cosmology, P. O. Box 2450, Stanford University, Stanford, CA 94305, USA}
\affiliation{SLAC National Accelerator Laboratory, Menlo Park, CA 94025, USA}

\author{Robert Gruendl}
\affiliation{Department of Astronomy, University of Illinois at Urbana-Champaign, Urbana, IL 61801, USA}
\affiliation{National Center for Supercomputing Applications, University of Illinois at Urbana-Champaign, Urbana, IL 61801, USA}

\author[0000-0003-3023-8362]{Julia Gschwend}
\affiliation{Laborat\'orio Interinstitucional de e-Astronomia - LIneA, Rua Gal. Jos\'e Cristino 77, Rio de Janeiro, RJ - 20921-400, Brazil}
\affiliation{Observat\'orio Nacional, Rua Gal. Jos\'e Cristino 77, Rio de Janeiro, RJ - 20921-400, Brazil}

\author[0000-0003-0825-0517]{Gaston Gutierrez}
\affiliation{Fermi National Accelerator Laboratory, P. O. Box 500, Batavia, IL 60510, USA}

\author{Samuel Hinton}
\affiliation{School of Mathematics and Physics, University of Queensland,  Brisbane, QLD 4072, Australia}

\author{Devon L. Hollowood}
\affiliation{Santa Cruz Institute for Particle Physics, Santa Cruz, CA 95064, USA}

\author[0000-0002-6550-2023]{Klaus Honscheid}
\affiliation{Center for Cosmology and Astro-Particle Physics, The Ohio State University, Columbus, OH 43210, USA}
\affiliation{Department of Physics, The Ohio State University, Columbus, OH 43210, USA}

\author[0000-0003-2511-0946]{Nikolay Kuropatkin}
\affiliation{Fermi National Accelerator Laboratory, P. O. Box 500, Batavia, IL 60510, USA}

\author[0000-0001-9856-9307]{Marcio Maia}
\affiliation{Laborat\'orio Interinstitucional de e-Astronomia - LIneA, Rua Gal. Jos\'e Cristino 77, Rio de Janeiro, RJ - 20921-400, Brazil}
\affiliation{Observat\'orio Nacional, Rua Gal. Jos\'e Cristino 77, Rio de Janeiro, RJ - 20921-400, Brazil}

\author{Marisa March}
\affiliation{Department of Physics and Astronomy, University of Pennsylvania, Philadelphia, PA 19104, USA}

\author[0000-0003-0710-9474]{Jennifer Marshall}
\affiliation{George P. and Cynthia Woods Mitchell Institute for Fundamental Physics and Astronomy, and Department of Physics and Astronomy, Texas A\&M University, College Station, TX 77843,  USA}

\author[0000-0002-4279-4182]{Paul Martini}
\affiliation{Center for Cosmology and Astro-Particle Physics, The Ohio State University, Columbus, OH 43210, USA}
\affiliation{Department of Physics, The Ohio State University, Columbus, OH 43210, USA}

\author[0000-0002-8873-5065]{Peter Melchior}
\affiliation{Department of Astrophysical Sciences, Princeton University, Peyton Hall, Princeton, NJ 08544, USA}

\author[0000-0002-1372-2534]{Felipe Menanteau}
\affiliation{Department of Astronomy, University of Illinois at Urbana-Champaign, Urbana, IL 61801, USA}
\affiliation{National Center for Supercomputing Applications, University of Illinois at Urbana-Champaign, Urbana, IL 61801, USA}

\author[0000-0002-6610-4836]{Ramon Miquel}
\affiliation{Instituci\'o Catalana de Recerca i Estudis Avan\c{c}ats, E-08010 Barcelona, Spain}
\affiliation{Institut de F\'{\i}sica d'Altes Energies (IFAE), The Barcelona Institute of Science and Technology, Campus UAB, 08193 Bellaterra (Barcelona) Spain}

\author{Francisco Paz-Chinchon}
\affiliation{Institute of Astronomy, University of Cambridge, Madingley Road, Cambridge CB3 0HA, UK}
\affiliation{National Center for Supercomputing Applications, University of Illinois at Urbana-Champaign, Urbana, IL 61801, USA}

\author[0000-0002-2598-0514]{Andr\'es Plazas Malag\'on}
\affiliation{Department of Astrophysical Sciences, Princeton University, Peyton Hall, Princeton, NJ 08544, USA}

\author[0000-0002-9328-879X]{Kathy Romer}
\affiliation{Department of Physics and Astronomy, Pevensey Building, University of Sussex, Brighton, BN1 9QH, UK}

\author[0000-0002-9646-8198]{Eusebio Sanchez}
\affiliation{Centro de Investigaciones Energ\'eticas, Medioambientales y Tecnol\'ogicas (CIEMAT), Madrid, Spain}

\author{Vic Scarpine}
\affiliation{Fermi National Accelerator Laboratory, P. O. Box 500, Batavia, IL 60510, USA}

\author[0000-0001-9504-2059]{Michael Schubnell}
\affiliation{Department of Physics, University of Michigan, Ann Arbor, MI 48109, USA}

\author{Santiago Serrano}
\affiliation{Institut d'Estudis Espacials de Catalunya (IEEC), 08034 Barcelona, Spain}
\affiliation{Institute of Space Sciences (ICE, CSIC),  Campus UAB, Carrer de Can Magrans, s/n,  08193 Barcelona, Spain}

\author[0000-0002-1831-1953]{Ignacio Sevilla}
\affiliation{Centro de Investigaciones Energ\'eticas, Medioambientales y Tecnol\'ogicas (CIEMAT), Madrid, Spain}

\author[0000-0002-3321-1432]{Mathew Smith}
\affiliation{School of Physics and Astronomy, University of Southampton,  Southampton, SO17 1BJ, UK}

\author[0000-0002-7047-9358]{Eric Suchyta}
\affiliation{Computer Science and Mathematics Division, Oak Ridge National Laboratory, Oak Ridge, TN 37831}

\author[0000-0003-1704-0781]{Gregory Tarle}
\affiliation{Department of Physics, University of Michigan, Ann Arbor, MI 48109, USA}

\author{Tamas Norbert Varga}
\affiliation{Max Planck Institute for Extraterrestrial Physics, Giessenbachstrasse, 85748 Garching, Germany}
\affiliation{Universit\"ats-Sternwarte, Fakult\"at f\"ur Physik, Ludwig-Maximilians Universit\"at M\"unchen, Scheinerstr. 1, 81679 M\"unchen, Germany}

\author{Reese Wilkinson}
\affiliation{Department of Physics and Astronomy, Pevensey Building, University of Sussex, Brighton, BN1 9QH, UK}

\begin{abstract}
The size of the dust torus in Active Galactic Nuclei (AGN) and their high-luminosity counterparts, quasars, can be inferred from the time delay between UV/optical accretion disk continuum variability and the response in the mid-infrared (MIR) torus emission. This dust reverberation mapping (RM) technique has been successfully applied to $\sim 70$ $z\lesssim 0.3$ AGN and quasars. Here we present first results of our dust RM program for distant quasars covered in the SDSS Stripe 82 region combining $\sim 20$-yr ground-based optical light curves with 10-yr MIR light curves from the WISE satellite. We measure a high-fidelity lag between W1-band (3.4~\micron) and $g$ band for 587 quasars over $0.3\lesssim z\lesssim 2$ ($\left<z\right>\sim 0.8$) and two orders of magnitude in quasar luminosity. They tightly follow (intrinsic scatter $\sim 0.17$\,dex in lag) the IR lag-luminosity relation observed for $z<0.3$ AGN, revealing a remarkable size-luminosity relation for the dust torus over more than four decades in AGN luminosity, with little dependence on additional quasar properties such as Eddington ratio and variability amplitude. This study motivates further investigations in the utility of dust RM for cosmology, and strongly endorses a compelling science case for the combined 10-yr Vera C. Rubin Observatory Legacy Survey of Space and Time (optical) and 5-yr Nancy Grace Roman Space Telescope $2$\micron\ light curves in a deep survey for low-redshift AGN dust RM with much lower luminosities and shorter, measurable IR lags. The compiled optical and MIR light curves for 7,384 quasars in our parent sample are made public with this work.   

\end{abstract}


\section{Introduction} \label{sec:introduction}
In the widely accepted unified model of AGN\footnote{In this paper AGN refer to SMBHs that are accreting efficiently, with a standard optically-thick, geometrically-thin accretion disk \citep{Shakura_Sunyaev_1973}. }, the inner regions surrounding the accreting supermassive black hole (SMBH) include (roughly from inside out but with potential spatial overlaps): a hot X-ray-emitting corona, an accretion disk, a broad emission-line region, a dusty toroidal structure, and a narrow emission-line region. The toroidal dusty region, commonly referred to as the dust torus, plays the central role in the AGN unification scheme \citep{Antonucci1993, Urry1995} that unifies type 1 (unobscured, broad-line) and type 2 (obscured, narrow-line) AGN. 

\begin{figure*}[!ht]
\centering
\includegraphics[width=0.9\textwidth]{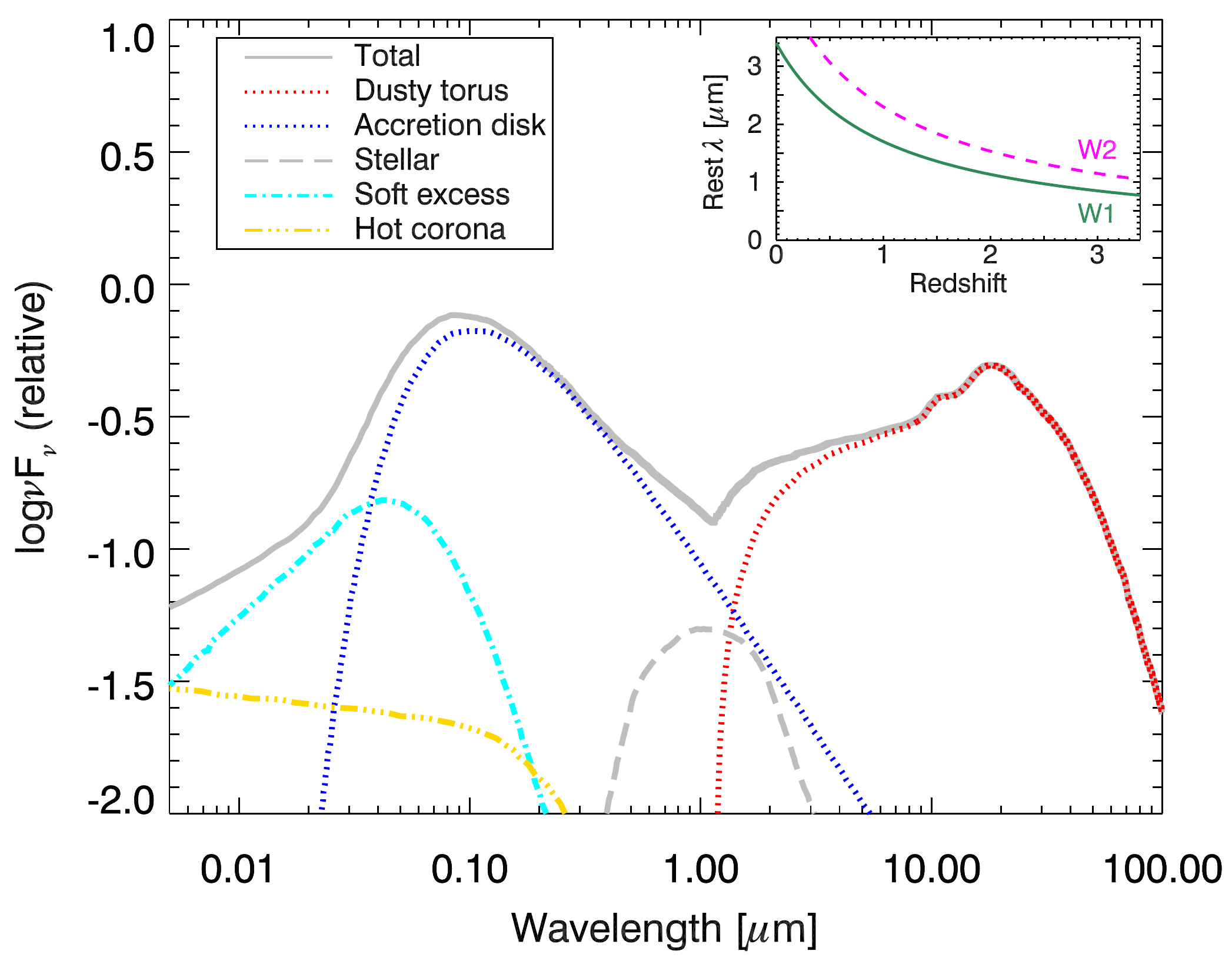}
\caption{A schematic representation of the broad-band SED of an unobscured broad-line quasar, including contributions from the accretion disk \citep{Shakura_Sunyaev_1973}, dusty torus \citep[e.g.,][]{Honig2006, Nenkova2008}, stellar emission \citep{Bruzual_Charlot_2003}, hot corona \citep{Haardt1991}, and soft excess (e.g., \citealt{Done2012} and references there-in).
The upper-right inset shows the rest-frame wavelengths in W1 and W2 bands as a function of redshift. This schematic is for demonstration purposes and the relative contributions are not exact and vary from object to object. 
\label{fig:SED}}
\end{figure*}

The spectral energy distributions (SEDs) of AGN show a pronounced peak in the mid-infrared \citep[MIR; e.g.,][]{Sanders1989, Elvis1994}, which is interpreted as thermal emission from hot dust in the torus region. The dust grains in the torus absorb the incident UV/optical continuum emission from the accretion disk and re-radiate in the infrared \citep{Rieke1978}, with a time lag corresponding to the average light travel time from the accretion disk to the dust torus. The torus extends from the dust sublimation radius outwards \citep{Barvainis1987}, with the near-infrared (NIR) radiation arising at the inner edge near the sublimation radius, and longer wavelength radiation from the outer regions with lower dust temperatures \citep[e.g.,][and references therein]{Nenkova2008,Netzer_2015}. Figure \ref{fig:SED} shows a schematic for the broad-band SED of unobscured broad-line AGN. 

The compact size ($\sim$sub-parsec to parsec for typical quasars) of the torus is difficult to be spatially resolved directly. There are only a handful of nearby bright AGN for which we can marginally resolve the torus structure using IR interferometric techniques \citep[e.g.,][]{Swain2003, Jaffe2004, Tristram2007, Kishimoto2009, Kishimoto2011a, Weigelt2012, Burtscher2013, GRAVITY2019}. Beyond the nearby Universe it becomes difficult for IR interferometry to directly resolve the torus due to the reduced angular size and brightness of the distant source. 

Dust reverberation mapping (RM) offers an alternative route to infer the size of the AGN torus without the need for spatial resolution. The echo of reprocessed dust emission to UV/optical continuum variations measures the average light-crossing time (hence a typical size) of the dust torus to the central engine. The dust RM lag has been measured between the UV/optical and the NIR \citep[e.g.,][]{Penston1971, Clavel1989, Glass1992, Glass2004, Oknyanskii1993, Oknyanskij1999, Nelson1996, Minezaki2004, Suganuma2006, Koshida2009, Koshida2014, Minezaki2019}, or UV/optical versus MIR \citep{Vazquez2015, Lyu2019}. These previous dust RM studies have confirmed correlated variability between the optical and NIR ($K$ band) emission, justifying the RM technique. The dust torus radius, $R$, inferred from the lag between the optical and NIR variability, was found to scale with the AGN luminosity, $L$, as $R \propto L^{1/2}$ \citep[e.g.,][]{Suganuma2006, Koshida2014}. Comparisons between the dust RM lags and direct interferometric measurements found general agreement between the size measurements from the two methods \citep[e.g.,][]{Kishimoto2011a}.

To date, most of the dust RM works on torus size measurements were limited to relatively low redshift ($z<0.3$). It is important to extend dust RM to high-redshift quasars to better sample the high-luminosity end of the $R-L$ relation, and to investigate any redshift evolution in the physical properties (such as size and grain physics) of dust torus. At higher redshifts, we can only observe luminous quasars, whose observed-frame torus lags are longer than their low-$z$ counterparts both due to the higher luminosity and cosmic time dilation. Therefore a sufficiently long time baseline is required to measure the dust RM lag for these distant and luminous quasars.   

In this paper, we present dust RM measurements for a large sample of quasars at a median redshift of $\left<z\right>\sim 0.8$. We make use of the multi-epoch MIR imaging from the all-sky {\it Wide-field Infrared Survey Explorer} \citep[{\it WISE},][]{Wright2010, Mainzer2014}, combined with optical light curves from all available ground-based facilities, including multi-epoch data from the Dark Energy Survey \citep[DES,][]{Abbott2018}, SDSS \citep{York2000}, Pan-STARRS \citep[PS1,][]{Chambers2016}, the Catalina Real-time Transient Survey \citep[CRTS,][]{Drake2009}, the Palomar Transient Factory \citep[PTF;][]{Law2009}, the All-Sky Automated Survey for Supernovae \citep[ASAS-SN,][]{Shappee_etal_2014}, and the Zwicky Transient Facility \citep[ZTF,][]{Bellm2019}. The WISE MIR data covers a baseline of $\sim 10$ yrs with a cadence of $\sim$ 6 months, and the optical data covers a combined baseline of $\sim$ 20 years. These multi-year baselines provide a unique opportunity to measure the dust echos in luminous and distant quasars. \citet{Lyu2019} already demonstrated the power of combining WISE light curves with ground-based optical light curves in measuring dust echos in luminous Palomar-Green (PG) quasars at $z<0.5$. Here we extend this exercise to even higher redshifts with SDSS quasars. At the median redshift of our sample ($\left<z\right>\sim 0.8$), WISE $W1$ (3.4 $\mu$m) and $W2$ (4.6 $\mu$m) data mainly probe the dust emission at rest-frame 2\,$\mu$m, allowing us to directly compare the MIR lags of these quasars with the NIR (mostly $K$ band) lags in nearby AGN \citep[e.g.,][]{Suganuma2006,Koshida2014}.

The structure of this paper is as follows. In \S\ref{sec:data}, we describe our quasar sample and the photometric data (in MIR and optical). We describe the dust RM measurements in \S\ref{sec:lag}. We present the results of MIR lags in \S\ref{sec:results}, and discuss the relation between the dust radius and quasar luminosity in \S\ref{sec:disc}. We conclude in \S\ref{sec:summary} with an outlook for future work. Throughout this paper, we adopt a flat $\Lambda$CDM cosmology with parameters $\Omega_{\Lambda}=0.7$, $\Omega_{\rm m}=0.3$, and $H_0=70$ km s$^{-1}$ Mpc$^{-1}$. All quoted uncertainties are 1$\sigma$.

\section{Sample and Data} \label{sec:data}

\begin{deluxetable*}{lcccccc}
\tablecaption{Survey Information \label{tab:Optical}}
\tablewidth{1pt}
\tablehead{
\colhead{Survey} &
\colhead{Filter} &
\colhead{Time} &
\colhead{Cadence} &
\colhead{N$_{\rm epoch}$} &
\colhead{Coverage} &
\colhead{Depth}
}
\startdata
SDSS(S82) & $g$ & 1998-2007 & 5 days & $\sim$60 & 300 deg$^2$ & 22.2 \\
PS1 & $g$ ($r$, $i$) & 2011-2014 & $2$/season &$\sim$10  & 3$\pi$ & 22.0\\
DES & $g$ ($r$, $i$) & 2013-2018 &  1-4/season (wide-field) & $\sim$10 & 5100 deg$^2$ & 23.57 \\
PTF & $g$ & 2009–2014 & 5 days& $\sim$4 & 11,233 deg$^2$ & 19 \\
ZTF & $g$ & 2018 & 3 days & $\sim$30 & 2.5-3$\pi$ & 20.5 \\
CRTS & unfiltered & 2005–2013 & 3\,weeks & $\sim$40 & 33,000 deg$^2$ & 19-21 \\
ASAS-SN & V & 2012–2019 & 3-4 days & $\sim$200 & all sky & 17 \\
WISE & $W1$ ($W2$) & 2010–2019 & 6 months & 12-15 & all sky & 17.6 (Vega) \\
\enddata
\tablecomments{The cadence of WISE light curves is 6 months per visit. Optical data were obtained in annual ``seasons". For example, SDSS S82 data were obtained annually within a 2–3 months window and the cadence effectively samples timescales from days to years. For a single season, the median SDSS cadence is $\sim$ 5 days. Since most quasars in our sample are fainter than 19 mag in $r$ band, we bin the CRTS and PTF data annually (inverse-variance weighted mean) to increase the signal-to-noise ratio. The exact cadences in most of these optical surveys are more complicated than the quoted approximate values, given various observing constraints. }
\end{deluxetable*}

\subsection{The S82 Quasar sample} \label{sec:sample}

Our parent quasar sample includes the 9,258 spectroscopically confirmed broad-line quasars in the Stripe 82 (S82) region \citep{MacLeod_etal_2012} that are included in the SDSS DR7 quasar catalog \citep{Shen2011}. 
There are several advantages of this sample for our dust RM study for high-$z$ quasars: (1) these quasars cover a broad range in redshift and quasar luminosity, and are representative of the luminous quasar population at high redshift; (2) they are bright enough for reliable photometric measurements from ground-based imaging surveys with small-aperture telescopes and from WISE; (3) they have $\sim 20$ years of optical light curves combining all available photometry from various surveys described in \S\ref{sec:introduction}; (4) they have well measured spectroscopic properties, such as BH mass and Eddington ratios ($L/L_{\rm Edd}$) from the \citet{Shen2011} catalog.

Our targets well sample the high-luminosity regime in the torus size-luminosity relation, compared with earlier dust RM measurements in low-redshift AGN and PG quasars \citep[e.g.,][]{Suganuma2006,Koshida2014,Lyu2019}. Many of these S82 quasars have observed-frame MIR lags more than a few years, therefore it is necessary to have decade-long light curves to meaningfully measure the lag. There are still selection biases for the highest redshift/luminosity quasars due to the duration of our light curves, which will be discussed in \S\ref{sec:disc}.

\subsection{{\it WISE} Light Curves}\label{sec:wise}

{\it WISE} scanned the full sky from January to July in 2010 in four bands centered at wavelengths of 3.4, 4.6, 12, and 22 $\mu$m ($W1$, $W2$, $W3$, and $W4$). The secondary cryogen survey and Near-Earth Object Wide-field Infrared Survey Explorer \citep[{\it NEOWISE};][]{Mainzer2011} Post-Cryogenic Mission mapped the sky from August, 2010 to February, 2011. The {\it NEOWISE} Reactivation Mission \citep[{\it NEOWISE-R};][]{Mainzer2014} surveys the sky in $W1$ and $W2$ bands from 2013 twice a year. {\it WISE} obtains $\sim 10-20$ observations within a 36-hrs window in each visit. We calculate the median magnitude and magnitude error, specifically the semi-amplitude of the range enclosing the 16th and 84th percentiles of all flux measurements within a 6-month window. We limit to good quality single-epoch data points with the best frame image quality score ($qi\_fact=1$), observed far away from the South Atlantic Anomaly ($saa\_sep \geq 5$), with no contamination from the moon ($moon\_masked=0$), and excluding spurious detection ($cc\_flags=0$). The {\it WISE} magnitudes are profile-fitting magnitudes, and are converted from Vega to AB magnitude as $m_{\rm AB} = m_{\rm Vega} + \Delta m$, where $\Delta m$ is 2.699, 3.339, 5.174, and 6.620 in $W1$, $W2$, $W3$, and $W4$ bands, respectively \citep{Jarrett2011}.

\begin{table}  
\caption{Sample Statistics} \label{tab:sample}
\begin{center}
\begin{tabular}{lcc} 
\hline \hline
Cut & Number & Section \\
\hline
all & 9,258 & \S\ref{sec:sample} \\
.. matched in WISE & 8,990  & \S\ref{sec:wise} \\
.. covered in DES & 7,384 & \S\ref{sec:opt} \\
lag quality cuts & & \\
.... ${\rm SNR_{MIR}}>4$ & 1,315 & \S\ref{sec:mir_var}\\
.... autocut & 1,064 &  \S\ref{sec:finalcut}\\
.... finalcut & 587 & \S\ref{sec:finalcut}\\
\hline
\end{tabular}
\end{center}
\end{table}

We extract the WISE light curves from the latest observations up to December 13, 2019 (released on March 26, 2020) for 
$9,258$ S82 quasars, using a matching radius of 2\arcsec. Since W1 is the most sensitive band with the highest fraction of WISE detection of S82 quasars, we focus on the W1-band MIR light curves in this work. The results of multi-band (W1 and W2) MIR lags and implications on dust torus properties (such as the radial temperature profile) will be presented in a future paper.  

\subsection{Optical Light Curves}\label{sec:opt}

We compile all available optical photometric data from various ground-based imaging surveys that cover the S82 region, including SDSS, PS1, DES, CRTS, ASAS-SN, PTF, and ZTF (see Table \ref{tab:Optical}). These ground-based optical surveys cover different epochs and have different bandpasses and depths. We homogenize these optical light curves as described in detail in \S\ref{sec:opt_calib}. Table \ref{tab:Optical} summarizes the characteristics of these optical surveys. The default optical magnitude type is the PSF magnitude. 

Since WISE spans from 2010 to 2019, DES data (covering $\sim 2013-2018$) are crucial for measuring the MIR time lag. We therefore restrict our analysis to the subset of 7,582 quasars from the S82 sample located at ${\rm RA}<46$\,deg or ${\rm RA}>316$\,deg within the footprint of the DES wide survey. 7,384 of these quasars have available WISE light curves in W1. We summarize the sample statistics in Table \ref{tab:sample}.

\subsection{Optical Photometric Calibration}\label{sec:opt_calib}

\begin{deluxetable*}{lccl}
\tablecaption{FITS Table Format for Compiled Light Curves \label{tab:LC}}
\tablewidth{1pt}
\tablehead{
\colhead{Column} &
\colhead{Format} &
\colhead{Units} &
\colhead{Description}
}
\startdata
DBID	& LONG & & Object ID of SDSS S82 quasars \\       
MJD & DOUBLE & days & Modified Julian Date \\
SURVEY & STRING & & Name of the imaging survey \\
BAND	& STRING	& & Photometric filter \\
MAG & DOUBLE & mag & Optical magnitude in AB; WISE magnitude in Vega \\
MAG\_ERR & DOUBLE & mag & Uncertainty in magnitude \\
FLUX & DOUBLE & mJy & Flux density\\
FLUX\_ERR & DOUBLE & mJy & Uncertainty in flux density \\
\enddata
\tablecomments{This table compiles light curves for all 7,348 S82 quasars with both DES and WISE coverage. Each row corresponds to a single epoch in a given survey and filter for a single object. All optical magnitudes and fluxes are converted to DES $g$ band as described in \S\ref{sec:opt_calib}.}
\end{deluxetable*}

To calibrate the optical data from different surveys onto the same flux scale, we apply additive corrections to the optical magnitudes taking into account differences in filter curves and reported photometric magnitudes. To correct for different filter curves, we convolve the SDSS spectrum with the PS1/DES/ZTF filter curves to obtain synthetic magnitudes, and compare to those derived with the SDSS filters to derive the corrections. CRTS data are observed through an unfiltered wide band, so we apply a constant offset to the CRTS magnitudes to match the median CRTS magnitude to the contemporary calibrated PS1 magnitudes.

All optical data are then cross-calibrated to DES $g$-band, and converted to physical fluxes for our lag measurements. For surveys with multi-band coverage, we also include $r$-band data (converted to $g$ band) to increase the cadence; we only include PS1 and DES $i$-band data (converted to $g$ band) when there is no $g$-band or $r$-band data within $\pm1$ yr; no other bands were used given uncertainties in color transformations. We can safely ignore the small delays ($\lesssim$ a few days) across optical continuum bands due to accretion disk RM for our quasars \citep[e.g.,][]{Jiang_etal_2017,Mudd_etal_2018,Homayouni_etal_2019,Yu_etal_2020a}. 

In this work we do not correct for host galaxy contamination in the photometry since the vast majority of our quasars are at $z>0.5$ and are luminous enough to dominate the emission in both optical and MIR. Constant host stellar emission also does not affect lag measurements. The intrinsic MIR variability required for successful lag detection (see \S\ref{sec:finalcut}) is significantly higher than any systematic magnitude uncertainties due to host galaxy photometry.\footnote{To investigate the stability of WISE epoch-by-epoch photometry for galaxies, we use $\sim 15,000$ $z<0.3$ SDSS star-forming galaxies within the S82 region. The measurement uncertainty-subtracted, median RMS for these galaxies is 0.02 mag (0.0006 mag) in W1 (W2). This systematic uncertainty is far too small to make any impact on our lag measurements, where the quasar light largely dominates over host galaxy light in W1. A small fraction ($0.8\%$) of these galaxies do show significant intrinsic variability, many of which are due to obscured type 2 AGN or transient events \citep[see a dramatic example in, e.g.,][]{Yang_etal_2019b}. } Even in the worst case scenario where the epoch-by-epoch host galaxy photometry introduces significant, random variability in the light curves, it will only make the lag more difficult to detect rather than bias the lag in any particular direction. We also do not consider the contamination of broad emission line flux in the broad photometric bandpasses. The delayed response in broad-line flux to optical continuum variations occurs on much shorter light-crossing time than the torus lag, and therefore neglecting this complication will not affect the MIR lag measurements. 

The final merged optical light curves and WISE W1 light curves for all 7,384 quasars are available through the electronic version of this paper as an online FITS table. The format of the table is described in Table \ref{tab:LC}. While these light curves have the longest duration compared to data sets used in earlier studies, their cadences are typically insufficient for BLR RM. Fortunately for dust RM, the much extended torus entails much longer time delays and broader transfer functions, which makes MIR lag measurements possible even with the sparse sampling of the WISE light curves. 

\section{Lag Measurements}\label{sec:lag}

Robustly measuring the lag between two sets of light curves is a non-trivial task, which depends on the quality of the light curves (e.g., duration, cadence, signal-to-noise ratio) as well as the intrinsic variability of the light curves. The success of the lag measurement critically depends on the prominent features in the variable light curve, thus only for quasars with significant variability during the monitoring period can we measure a reliable time lag. Furthermore, quasar variability is stochastic in nature, and given insufficient time baselines or cadences, analyses of the light curves can produce artificial lags (``aliases''), often reaching the edges of the light curves where there is limited overlap in data points of the pair of light curves. 

There are various techniques to measure the time lag between two light curves. The most commonly adopted method is the Interpolated Cross-Correlation Function \citep[ICCF,][see \S\ref{sec:ICCF}]{Gaskell1987, Peterson1998}. The major advantage of ICCF is that it is an empirical and fast method that is model-independent, and the simple linear interpolation scheme across light curve gaps can recover some variability information lost. There have been some recent modifications on the ICCF method \citep[e.g.,][]{Grier2017,Grier2019,Li_etal_2019} that incorporates a weighting scheme in ICCF to down-weight time delays with less overlapping data points in the light curves in the lag search. This modified method (WCCF, see \S\ref{sec:wccf}) proves to be effective in eliminating aliases near the edge of the light curves. 

A more recently-developed technique, implemented in the public code \javelin\citep{Zu_etal_2011}, improves the interpolation scheme within light curve gaps by assuming a damped random walk (DRW) model for stochastic variability of AGN. The DRW model has proven to be a reasonably good prescription to describe the optical continuum variability of quasars \citep[e.g.,][]{Kelly2009, Kozlowski2010, MacLeod2010, Zu2013} on timescales of interests to most RM studies of quasars, offering a superior interpolation scheme for the light curves than ICCF. \javelin also implements more statistically rigorous procedures to estimate the uncertainties in the interpolations and measurements of the lags with Bayesian inference. There are other public codes such as {\tt CREAM} \citep{Starkey_etal_2016} that model the driving light curve with alternative statistical methods and perform equally well as \javelin.

\begin{figure*}[htbp]
 \centering
 \hspace{0cm}
 \subfigure{
 \hspace{-1.0cm}
  \includegraphics[width=3.5in]{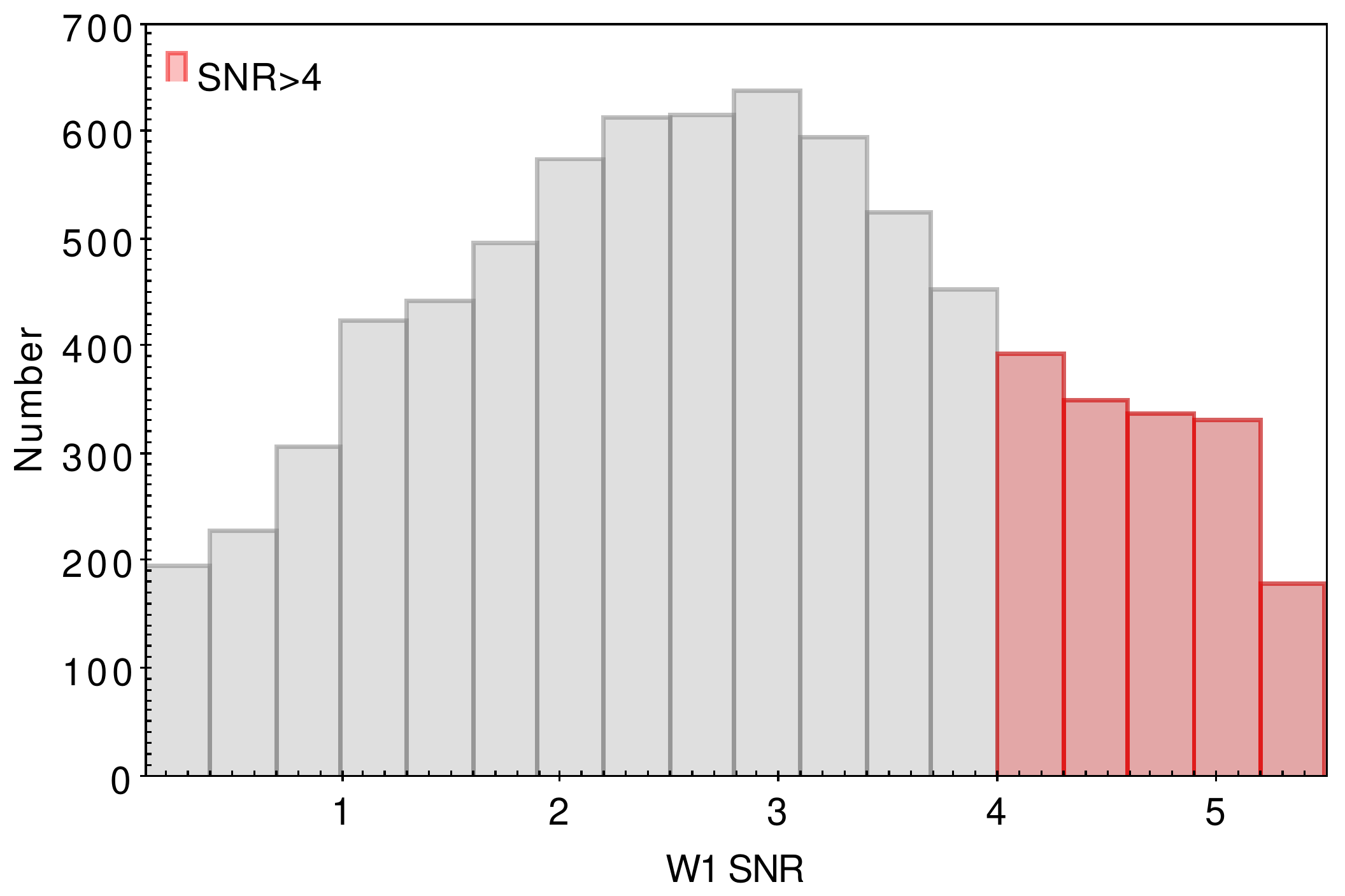}}
 \hspace{0cm}
 \subfigure{
  \includegraphics[width=3.5in]{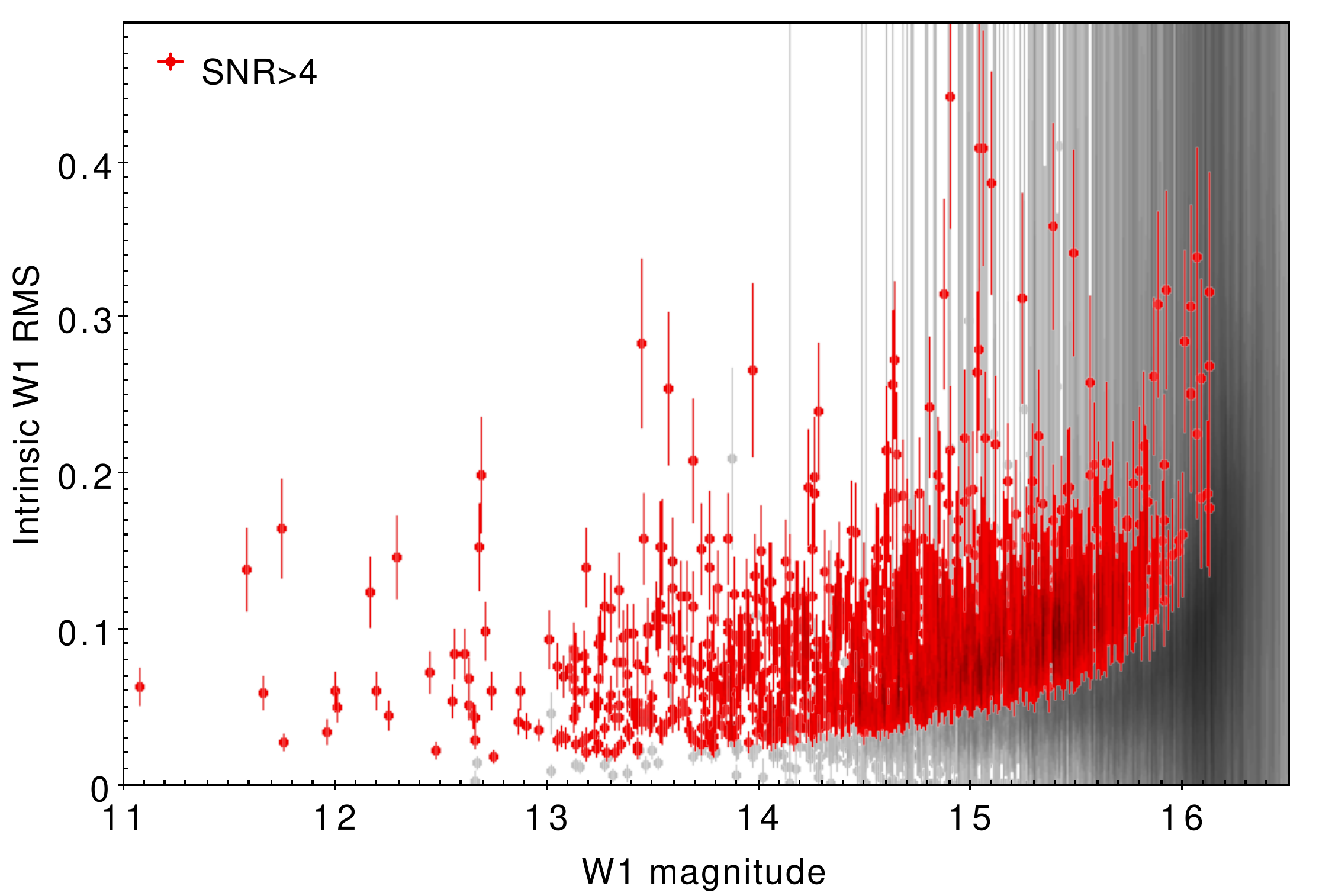}}\\
  \caption{{\it Left:} the histogram of the ${\rm SNR_{MIR}}$ for the intrinsic rms magnitude in W1 calculated for the light curve. {\em Right:} the distribution in the W1 intrinsic rms versus average W1 magnitude space for the parent sample (gray points) and the ${\rm SNR_{MIR}}>4$ subsample (red points) to illustrate the amplitudes and uncertainties of the intrinsic rms variability measurements. The intrinsic rms magnitude and its uncertainties are described in \S\ref{sec:mir_var}. \label{fig:SNR} }
\end{figure*}

\citet{Li_etal_2019} performed a detailed comparison of lag measurement methodologies for ICCF, \javelin, and the z-transformed Discrete Correlation Function \citep[zDCF,][]{Alexander_2013}, in particular for light curves with moderate-to-low qualities. They found that \javelin provides overall the best performance in terms of recovering the correct lag and estimating the lag uncertainties, while ICCF is less effective for low-to-moderate light curve quality, with zDCF being the least effective method. When the light curve quality is sufficiently high, all methods converge to consistent lag measurements. 

Since the light curve quality for our quasars is moderate-to-low, we will rely on the more robust lag measurement method \javelin\ to provide our fiducial measurements. However, we will also use the WCCF method to guide the \javelin\ measurements, and to provide additional criteria in eliminating false positives. We note that currently there is no perfect method to measure lags for light curves of different qualities. Thus we will also impose a set of cuts assisted by visual inspection to select a final ``cleaned'' lag sample (\S\ref{sec:finalcut}), as often done in recent studies \citep[e.g.,][]{Grier2017,Grier2019,Lyu2019}. 

\subsection{Intrinsic Variability Cut}\label{sec:mir_var}

To robustly detect the lag, the variability characteristics of the light curves are of critical importance. We define a variability metric to ensure variability is well detected in the MIR light curve. The observed rms variability includes both intrinsic variability and photometric uncertainties. To estimate the intrinsic rms magnitude (or ``excess rms'') of a light curve and the uncertainty of the intrinsic rms, we utilize a maximum-likelihood estimator detailed in \citet{Shen_etal_2019b} (their Equations 5--9)\footnote{We correct a typo in their Eqn. (9): ${\rm Var}[\mu]=\sigma_0^2/\Sigma g_i$.}. The estimate of the intrinsic rms in the MIR light curve, $\sigma_{\rm MIR}$, and its uncertainty $\Delta\sigma_{\rm MIR}$ are defined by Eqn.\ (8) of \citet{Shen_etal_2019b}. We then define the signal-to-noise ratio of the estimated intrinsic rms as ${\rm SNR_{MIR}} = \sigma_{\rm MIR}/\Delta\sigma_{\rm MIR}$. We require significant variability detection in the MIR light curve as ${\rm SNR_{MIR}}>4$. Among the 8,990 quasars with WISE light curves, 1,588 quasars have ${\rm SNR_{MIR}}>4$ in W1 band. Among the 7,582 quasars covered by DES, 7,384 are matched in WISE, and 1,315 of them have ${\rm SNR_{MIR}}>4$. Figure \ref{fig:SNR} shows the distribution of ${\rm SNR_{MIR}}$ (left) and the intrinsic rms magnitude in W1 band versus the mean W1 magnitude. Requiring significant variability in the MIR light curve implies the optical light curve generally also has large variations. Many of these highly variable quasars fall into the category of ``Extreme Variability Quasars'' \citep[EVQ,][]{Rumbaugh2018} that have more than 1 magnitude maximum variations in $g$-band over multi-year timescales. These gradual, large-amplitude multi-year variability features greatly facilitate the measurement of the MIR lag, as already demonstrated with earlier WISE light curves \citep[e.g.,][]{Sheng_etal_2017}.

\subsection{Interpolated Cross-Correlation Function}\label{sec:ICCF}

The cross-correlation function (CCF) is commonly used to measure the time delay between two light curves. 
For two sets of variables (e.g., light curves) $\bm{X}$ and $\bm{Y}$, the Pearson correlation coefficient, $r$, can be calculated by\\
\begin{equation}
    r = \frac{\sum_{i=1}^N (x_i-\overline{x})(y_i-\overline{y})}{\sqrt{\sum_{i=1}^N (x_i-\overline{x})^2\sum_{i=1}^N (y_i-\overline{y})^2}}
\end{equation}
where $\overline{x}$ and $\overline{y}$ are the mean of $\bm{X}$  and $\bm{Y}$ respectively, $x_i$ and $y_i$ are the $i$-th members, and N is the number of ($x_i$, $y_i$) pairs. For two time series, CCF is the Pearson correlation coefficient as a function of the time displacement ($\tau$)\footnote{We adopt the convention that positive values of $\tau$ correspond to the MIR light curve lagging behind the optical light curve. } of one signal relative to the other. The time delay between the two time series can be determined by the time displacement with the maximum CCF. Furthermore, the maximum CCF coefficient, $r_{\rm max}$, is a useful value to evaluate the time lag significance; a larger $r_{\rm max}$ indicates the two time series are better correlated \citep[e.g.,][]{Grier2017}. For uneven-sampled light curves in essentially all RM studies, ICCF is deployed to linearly interpolate the shifted light curves. We used the {\tt PyCCF} code \citep{Sun2018} to perform the ICCF calculations. 

\subsection{Weighted Cross-correlation Function}\label{sec:wccf}

Due to the relative sparse time sampling and the limited baseline of the light curves, there are often multiple peaks in the ICCF. The peaks near the boundaries of the lag search limited by the duration of the light curve, and sometimes within large gaps in the light curves, are usually false positives due to the small number of overlapping data points in the light curve pair \citep[but see rare counterexamples in e.g.,][]{Shen_etal_2019c}. There are formal estimations of the statistical significance of the correlation that depend on both the correlation coefficient $r$ and the number of overlapping data points \citep[e.g.,][]{Bevington_1969,Shen_etal_2015a}. However, in practice this statistical significance is almost never used because of the noisy ICCF for light curves with low-to-moderate quality. To remedy for these aliasing peaks due to less-overlapped data, \citet{Grier2017, Grier2019} developed a quantitative scheme taking into account the number of overlapping data points in the ICCF lag search. They used a weight function defined as $P(\tau)=[N(\tau)/N(0)]^2$, where $N(\tau)$ is the number of overlapping epochs as a function of time delay $\tau$. The exponent is somewhat arbitrary as long as it is positive to down-weight time delays with fewer overlapping data points. 

\citet{Li_etal_2019} further tested the efficiency and robustness of this weighting scheme with simulated light curves that mimic the low-to-moderate quality of data from recent RM survey programs, but with known input lags. They found that for survey-quality light curves, it is essentially necessary to impose this weighting scheme because the raw ICCF will produce an overwhelmingly large number of aliasing peaks in the less overlapped regime, as well as lags with large uncertainties. The weighted CCF (WCCF) is a much more robust method to recover the true lag.

To account for flux uncertainties in each epoch, we adopt a modified weight function considering both the number of overlapping epochs and the flux uncertainty of each epoch: 
\begin{equation}
    P(\tau) = \frac{\sum_{j=1}^n (1/\sigma_j)^2}{\sum_{i=1}^N (1/\sigma_i)^2}
\end{equation}
where $\sigma$ is the single-epoch flux measurement error, $n$ is the number of overlapping epochs at time delay $\tau$, and $N$ is the maximum number of overlapping epochs. In the case of equal flux uncertainties, our definition of the weight function is equivalent to $N(\tau)/N(0)$. Therefore the suppression of correlations at large absolute values of $\tau$ is not as aggressive as the weight function used in \citet[][]{Grier2017}. 

In principle the weight function should be defined for the optical light curve ($P_{\rm OPT}$) and the MIR light curve ($P_{\rm MIR}$) separately, since both light curves have different lengths, and the surveys contributing to the optical light curves have different sampling densities. A combined weight function $P_{\rm OM}\equiv P_{\rm OPT}P_{\rm MIR}$ can then be defined to impose stronger suppression of correlations in the less overlapped regime. However, since the MIR light curve is less well sampled than the optical light curve and dominates the cross-correlation signal, we only use $P_{\rm MIR}$ to down-weight the original ICCF. In addition, the weights computed for the optical light curve ($P_{\rm OPT}$) are complicated by the very different sampling densities and flux uncertainties in various surveys contributing to the light curve. Using $P_{\rm OM}\equiv P_{\rm OPT}P_{\rm MIR}$ thus could drastically change the shape of the ICCF, which is more than necessary to mitigate lag aliases in the less-overlapped regime. However, in defining the maximum search range in \javelin as discussed in \S\ref{sec:javelin}, using the more stringent weights $P_{\rm OM}$ is more efficient for the fit to converge.    

Multiplying the ICCF by $P_{\rm MIR}(\tau)$, we obtain the weighted CCF (WCCF) as 
\begin{equation}
    r_{\rm WCCF}(\tau) = r_{\rm ICCF}(\tau) \times P_{\rm MIR}(\tau).
\end{equation}
The maximum of $r_{\rm WCCF}(\tau)$, $r_{\rm wmax}$, is a parameter similar to $r_{\rm max}$ that can be used to evaluate the time lag significance that incorporates information from the weight function.

Figure \ref{fig:CCF} shows an example of ICCF and WCCF of a quasar in our S82 sample, as well as the weight functions $P_{\rm MIR}$ and $P_{\rm OPT}$. The WCCF efficiently eliminated the fake peaks near the boundaries of the lag search window and enhanced the primary CCF peak. Given the benefits of the WCCF, we will use the WCCF and the weight functions defined here to guide our \javelin measurements and to refine our criteria of significant lag detection in \S\ref{sec:finalcut}. But we note that our fiducial lags are computed using the more robust \javelin\ method, and the WCCF is used only for cross-checks and to prevent \javelin\ lag searches near the boundaries of the light curves.

\subsection{Lag Measurements with \javelin}\label{sec:javelin}

\begin{figure*}[!ht]
\centering
\includegraphics[width=0.9\textwidth]{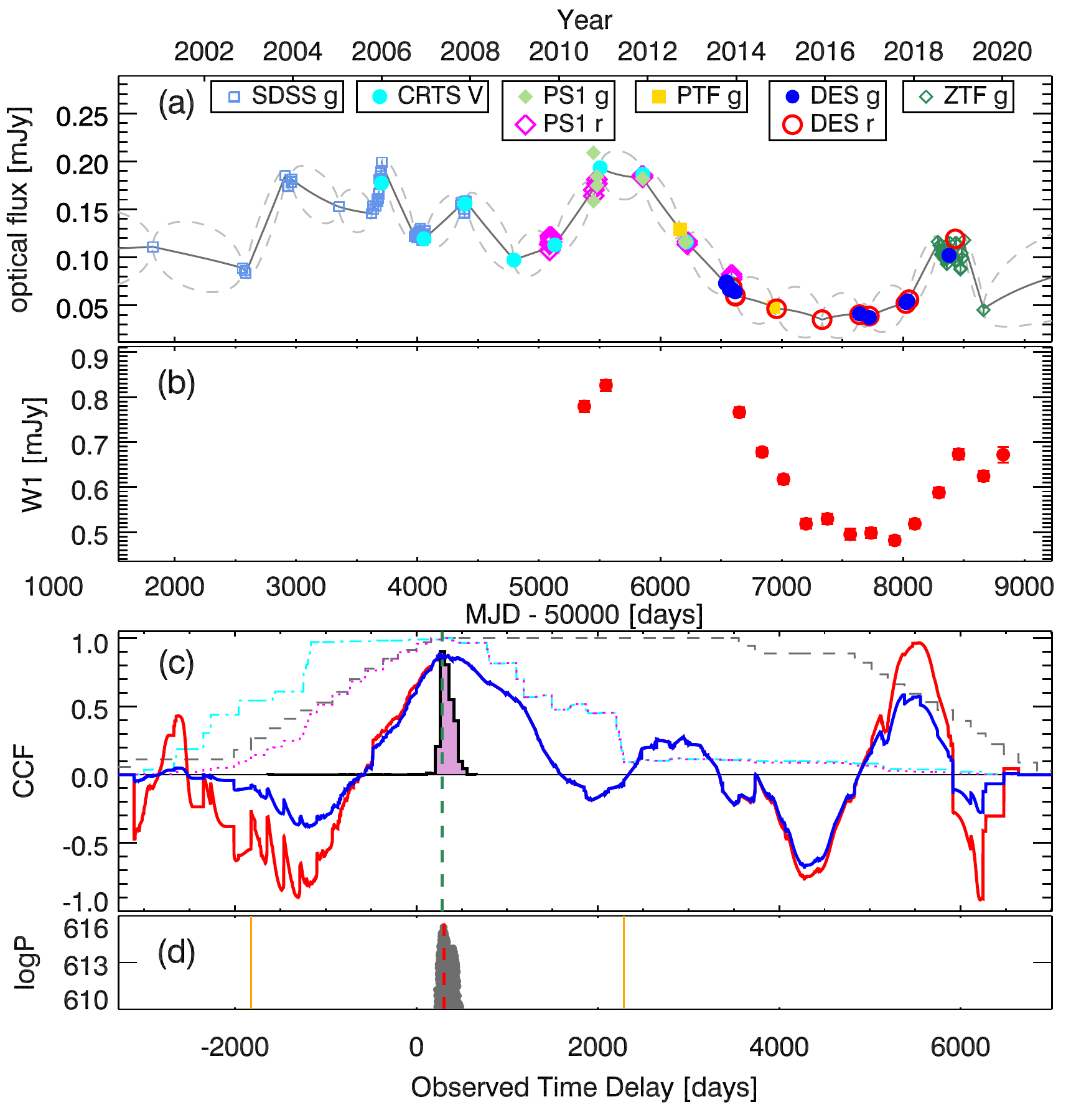}
\caption{An example quasar in our S82 sample, J0023+0035, at $z=0.4219$. The upper two panels show the light curves in optical (a) and MIR $W1$ band (b), respectively. We use combined optical data from multiple ground-based imaging surveys, including SDSS, CRTS, PS1, PTF, DES, and ZTF. 
In panel (c), the gray dashed line and the cyan dash-dotted line are the normalized weight functions from overlapped MIR and optical data ($P_{\rm MIR}(\tau)$ and $P_{\rm OPT}(\tau)$), respectively. The purple dotted line denotes $P_{\rm OM}$. The red solid line is the original ICCF and the blue solid line is the weighted ICCF (WCCF=ICCF$\times P_{\rm MIR}$). The WCCF efficiently eliminates aliasing peaks near the edges of the lag search range due to limited overlap in the light curves. The orchid histogram (with black solid outline) shows the \javelin\ posterior lag distribution in observed frame. The vertical green dash line shows the peak of the \javelin\ posterior distribution. We measure a MIR (W1) lag as $306.5^{+48.8}_{-25.8}$ days using \javelin.  Panel (d) shows the \javelin\ distribution of logarithmic probability versus lag. The red dashed line marks the lag with the maximum probability. The two orange vertical lines in panel (d) show the lag search window in \javelin\ defined by $P_{\rm OM}>0.1$. The consistency between the maximum-probability lag (panel d) and the peak of the lag posterior (panel c) is one of the criteria for our visual rejection of less secure lags (see \S\ref{sec:finalcut}).
\label{fig:CCF}}
\end{figure*}

In \javelin, the MIR light curve is a scaled and smoothed version of the driving optical/UV light curve due to the extended structure of the dust torus. The responding torus IR light curve is the convolution of the optical continuum light curve, $f_{\rm cont}(t)$, with a transfer function, $\Psi(\tau)$, determined by the geometry of the dust torus. At rest-frame $\sim 2$~\micron, the emission is dominated by the torus in most of our quasars. But in extreme cases, the accretion disk continuum may contribute a significant amount of IR emission, following the theoretical prediction of the $F_{\nu}\propto\nu^{1/3}$ law \citep{Kishimoto2008}, that is delayed relative to the optical light curve on much shorter timescales (the light-crossing time of the accretion disk) than the torus lag. 

Neglecting the IR emission from the accretion disk, the responding IR light curve, $f_{\rm IR}(t)$, can be written as\\
\begin{equation} \label{eq:1}
    f_{\rm IR}(t) = \alpha \int{d\tau \Psi(\tau)f_{\rm cont}(t-\tau)}
\end{equation}
where $\tau$ is the average light travel time from the accretion disk to the dust torus, and $\alpha$ is the ratio between the responding IR and optical variability amplitudes. 

For our purposes, we use the two-band photometric RM model in \javelin \citep{Zu_etal_2011}, which models the quasar continuum variability with the DRW model and measures the lag between two photometric bands as described in Equation \ref{eq:1}. For simplicity, we consider a top-hat transfer function for the dust torus:\\
\begin{equation}
    \Psi(t) = 
    \begin{cases}
    1, & t_1 \leq t \leq t_2\\
    0,              & \text{otherwise}
    \end{cases}
\end{equation}
which has a mean lag of $\overline{\tau}=(t_1+t_2)/2$ and a width of $\Delta \tau=(t_2-t_1)$. 
We allow \javelin to fit $\Delta\tau$ as a free parameter, and we find there is a broad trend that the average $\Delta\tau$ increases with the best-fit lag. 

For each of the quasars that pass the ${\rm SNR_{\rm MIR}}>4$ cut, we run \javelin\ using 50,000 Markov Chain Monte Carlo (MCMC) chains, sufficient for convergence. To reduce the most time-consuming chains near the boundaries of the time delays that can be reasonably ruled out by WCCF, we allow \javelin to explore a range of lags where $P_{\rm OM} > 0.1$ (typically $t_{\rm min} \sim -2000$ and $t_{\rm max} \sim 5500$ days). This maximum lag search range corresponds to $\sim 70\%$ of the maximum baseline defined by the optical or MIR light curve (or $\sim 30\%$ overlap in the optical+IR light curves), which is typically required as the bare minimum for robustly measuring a lag \citep[e.g.,][]{Grier2017,Grier2019}. In other words, the use of the WCCF weights does not impose stringent limits in the \javelin\ lag search other than eliminating lags (mostly false positives) near the boundaries of the light curves. 

In the example shown in Figure \ref{fig:CCF}, the black solid histogram in Panel (c) indicates the \javelin posterior distribution of lags. We obtain the median lag from the posterior as $\tau_{\javelin}$ and the $1\sigma$ lag uncertainty as the semi-amplitude of the range enclosing the 16th and 84th percentiles of the \javelin posterior distribution of the mean lag.

The moderate-to-low quality of our light curves generally does not allow us to explore more sophisticated torus transfer function forms, except for the best individual cases. However, we have tested varying the width of the top-hat transfer function, and confirm that this somewhat arbitrary choice of the transfer function form in JAVELIN does not affect much the measured mean lag and its uncertainties. This measured ``average'' lag simply reflects the inner boundary of the torus, as rest-frame $K$ band traces the hottest dust near the sublimation temperature \citep[e.g.,][]{Honig_Kishimoto_2011}.

\begin{figure*}[!ht]
\centering
\includegraphics[width=0.9\textwidth]{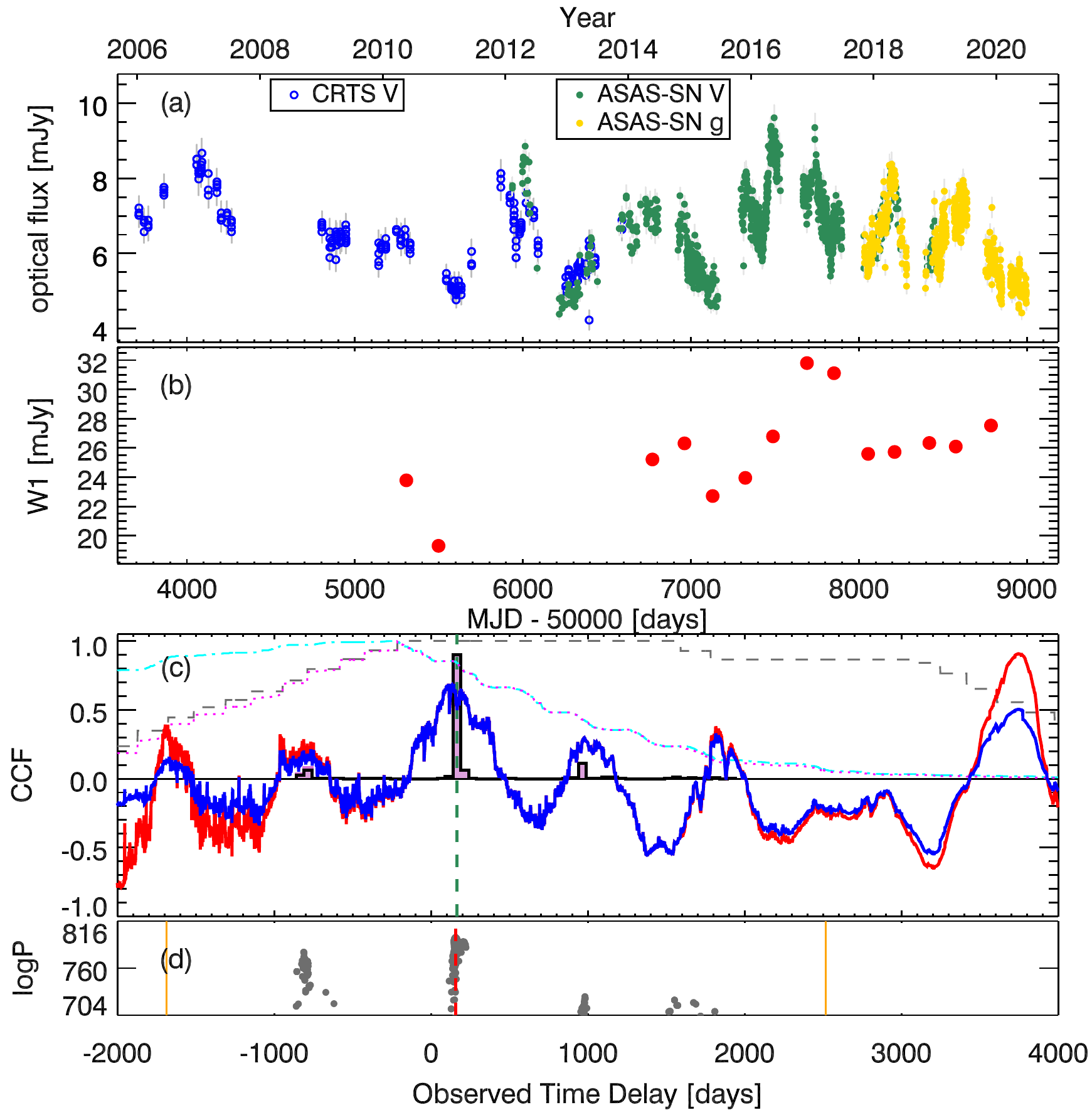}
\caption{An example of lag measurements using optical and WISE light curve for the nearby AGN MrK 110. Notations are the same as in Figure \ref{fig:CCF}. The optical fluxes are converted to ASAS-SN V band. \citet{Koshida2014} measured the K-band lag of Mrk 110 as $113.1^{+8.8}_{-8.6}$ and $124.1^{+7.1}_{-7.1}$ days in two campaigns in the observed frame. Using optical and WISE light curves, we successfully measure a MIR (W1) lag of Mrk 110 as $158.8^{+8.0}_{-6.4}$ days, where the longer lag is likely due to the longer wavelength IR band. The WISE light curve is apparently smoother than the optical light curve because of the averaging effect of responses from different parts of the extended torus.  
\label{fig:mrk110}}
\end{figure*}

To demonstrate the feasibility of our lag measurement approach, we show in Fig.~\ref{fig:mrk110} a local AGN (Mrk 110) with well measured K-band lags using earlier ground-based IR light curves. We successfully measured the lag with optical and WISE light curves, albeit with a slightly larger value due to the different (3.5\,\micron) IR band used. We have also applied our methodology to the entire PG quasar sample studied in \citet{Lyu2019} and found consistent results in general (see discussion in \S\ref{sec:others}), although the latter used a different method (a $\chi^2$ minimization scheme) to measure the lag.

\subsection{Criteria of Significant Lag Detection}\label{sec:finalcut}

To define a final ``cleaned'' sample of lags we perform the following automatic cuts and manual rejections with visual inspection. Most importantly, these rejection criteria do not use any prior information of an anticipated $R-L$ relation for the lags, therefore they do not introduce any bias to the observed $R-L$ relation.  

As described in \S\ref{sec:mir_var}, we first cut the sample (matched with WISE and DES) by ${\rm SNR_{MIR}}>4$, resulting in 1,315 quasars. We then require $r_{\rm wmax}>0.5$ in the WCCF (\S\ref{sec:wccf}) to remove objects where the optical and MIR light curves are not well correlated at any time lag within the entire lag search window. This step results in 1,283 quasars. To identify the primary peak of the lag distribution in the \javelin\ analysis, we bin the \javelin\ MCMC lag posterior with a bin size of 50 days. We identify the tallest peak of the binned distribution as the primary lag peak. From the primary peak, we search on each side to define the peak boundary where the binned number of lags falls below $5\%$ of the number of lags in the tallest bin. We require the fraction of lags within the primary peak to all lags $f_{\rm peak}>0.5$. This step using \javelin\ lag posterior is to ensure there is clear evidence for a primary peak in the lag distribution. There are 1064 quasars remaining after these automatic cuts (autocut). 

We then visually inspect the light curves, WCCF, and the \javelin\ lag posterior distribution in the remaining objects to remove unreliable lags (examples of each case are shown in the Appendix):
\begin{enumerate}
    \item We reject objects with large gaps ($\gtrsim 2$ yr) in the optical light curve, especially when the data do not overlap sufficiently to meaningfully measure the lag (see an example in Figure \ref{fig:gap}) -- 193 objects were removed. 
    
    \item We further require that the maximum-probability lag from the MCMC chains in \javelin\ is consistent within $1\sigma$ range of the primary peak of the lag posterior to ensure that the MCMC chains are well converged.  Otherwise, the MIR light curve might align with the optical light curve at more than one potential lags (see Figure \ref{fig:gap} for an example). In such cases, the MCMC chains are not well converged and the results are unreliable -- 171 objects with inconsistent primary peak and maximum probability lag were removed.
    
    \item Very rarely the MIR variability is dominated by one season (e.g., a flare or a systematic outlier). Figure \ref{fig:gap} shows an example of these objects that have extreme MIR variability in a single epoch and very noisy ICCF. We excluded six such objects.
    
    \item Finally, we remove objects with multiple peaks in the \javelin\ lag posterior, or one extreme broad peak spanning over thousands of days, or with an obvious secondary peak (see an example in Figure \ref{fig:gap}). This is to ensure that we only select objects with one well-defined primary lag peak, in assist to the autocut criterion of $f_{\rm peak}>0.5$ mentioned above. We removed 330 objects with multiple peaks. 
\end{enumerate}

Our finalcut sample includes 587 quasars. Although we do not require the lag to be positive, there are no negative lags remaining after our selection process.  

Fig.~\ref{fig:Lz} displays the distribution of the S82 parent sample and the finalcut 587 quasars with high-fidelity MIR lag measurements in the quasar luminosity-redshift plane. Objects are removed from the parent sample mainly due to the ${\rm SNR_{\rm MIR}}>4$ cut, as well as the difficulty to measure a long lag given the maximum baseline of the light curves. Nevertheless, the finalcut sample includes quasars over a broad range of redshift and luminosity that far extends the regime probed by previous samples with IR lag measurements at $z\lesssim 0.3$ with almost an order of magnitude increase in statistics. 

Fig.~\ref{fig:wccf_snr} displays the distribution of the parent quasar sample and the finalcut sample in the ${\rm SNR_{MIR}}$ and WCCF peak space. Such a plot is often used in recent large-scale broad-line RM programs to demonstrate the statistical detection of lags using light curves of low-to-moderate quality \citep[e.g.,][]{Shen_etal_2016a,Grier2017,Grier2019}. The lengths and overlap of our optical and MIR light curves allow negative (i.e., MIR leading optical) and positive lags in the approximate range of $[-3000,7000]$\,days. If there is statistical evidence of the MIR light curve lagging behind the optical light curve, we would observe an asymmetry in the lag distribution towards more positive lags. Examining the range of $\pm 3000$\,days in Fig.~\ref{fig:wccf_snr}, there is indeed a preference of positive lags, indicating that statistically the MIR light curve lags the optical, as expected from torus reprocessing. Fig.~\ref{fig:wccf_snr} suggests that there are potentially thousands of MIR lags that are real below the ${\rm SNR_{MIR}}=4$ cut. However, only the red points above the ${\rm SNR_{MIR}}=4$ cut provide the most secure individual lag measurements to study the relation between dust lags and quasar luminosity.  

We summarize our lag measurements for the 587 quasars in an online FITS table, and describe the columns in Table \ref{tab:FITS}. In the appendix (Figure \ref{fig:examples}) we show several additional examples of lag measurements in our finalcut sample. The full figure set for all these quasars is available at\\ {http://quasar.astro.illinois.edu/moutai/mir\_lag/}

\begin{deluxetable*}{lccl}
\tablecaption{FITS Table Format for the Finalcut Lag Sample \label{tab:FITS}}
\tablewidth{1pt}
\tablehead{
\colhead{Column} &
\colhead{Format} &
\colhead{Units} &
\colhead{Description}
}
\startdata
DBID	&	LONG		&			& Object ID of SDSS S82 quasars \\                              
RA		&	DOUBLE	&	degree	& J2000 R.A. \\ 
DEC		& DOUBLE 	& degree	& J2000 Decl.  \\ 
REDSHIFT	& DOUBLE	& 		& Spectroscopic redshift \\
LOGLBOL       & DOUBLE        & [\ergs]  & Bolometric luminosity from \citet{Shen2011}  \\ 
LOGLBOL\_ERR   & DOUBLE        & [\ergs]  &  Uncertainty in LOGLBOL \\
LOGBH   & DOUBLE      & [$M_\odot$]  & Fiducial single-epoch BH mass from \citet{Shen2011}\\
LOGBH\_ERR & DOUBLE   & [$M_\odot$] & Uncertainty in LOGBH\\
LOGEDD\_RATIO & DOUBLE &  & Eddington ratio based on fiducial single-epoch BH mass \\
N\_WISE & LONG & & Number of WISE epochs \\
W1\_AVG & DOUBLE & & W1-band weighted average magnitude \\
W1\_RMS & DOUBLE & & W1-band intrinsic RMS \\
W1\_SIGRMS & DOUBLE & & Uncertainty in W1\_RMS \\
W1\_SNR & DOUBLE & & S/N of W1-band intrinsic RMS \\
RMAX & DOUBLE & & Peak correlation $r_{\rm max}$ from ICCF \\
RWMAX & DOUBLE & & Peak correlation $r_{\rm wmax}$ from WCCF \\
PEAK\_WCCF & DOUBLE & days & Peak location of WCCF \\
F\_PEAK & DOUBLE & & $f_{\rm peak}$ (see \S\ref{sec:finalcut}) \\
TAU\_JAVELIN & DOUBLE & days & \javelin\ time delay in observed frame\\
TAU\_JAVELIN\_LOW & DOUBLE & days & 1$\sigma$ lower limit of TAU\_JAVELIN \\
TAU\_JAVELIN\_UPPER & DOUBLE & days & 1$\sigma$ upper limit of TAU\_JAVELIN \\
\enddata
\end{deluxetable*}

\subsection{Comparison with Earlier Work}\label{sec:others}

\begin{figure}
\centering
\includegraphics[width=0.5\textwidth]{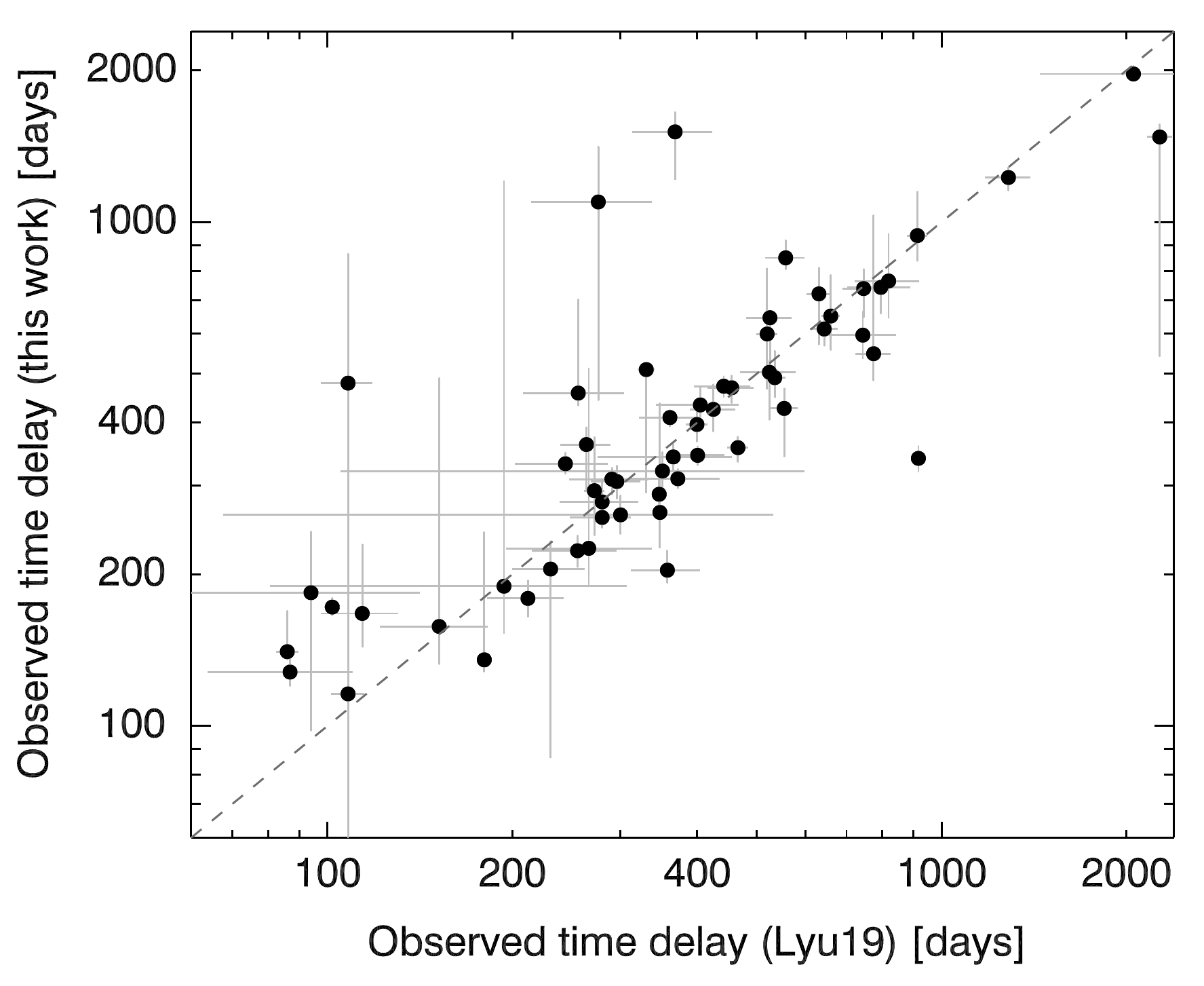}
\caption{Comparison of IR lag measurements for the PG quasar sample studied in \citet{Lyu2019} using optical and WISE-W1 light curves. Despite very different lag measurement methodologies, there is good agreement between the lags measured in both studies. The few outliers have low-quality WCCF ($r_{\rm max}<0.5$), so would not pass our selection criteria as reliable lags (\S\ref{sec:finalcut}). \label{fig:PG}}
\end{figure}

Using our lag measurement methodology, we re-analyze the PG quasar sample studied in \citet{Lyu2019} using our compiled optical and WISE light curves. Fig.~\ref{fig:PG} compares the lag measurements in both studies. We find good agreement between our measurements and those in \citet{Lyu2019}. This result confirms that our lag measurement methodology is robust, even though it differs significantly from the $\chi^2$-fit method adopted by \citet{Lyu2019}. 

\begin{figure}
\centering
\includegraphics[width=0.48\textwidth]{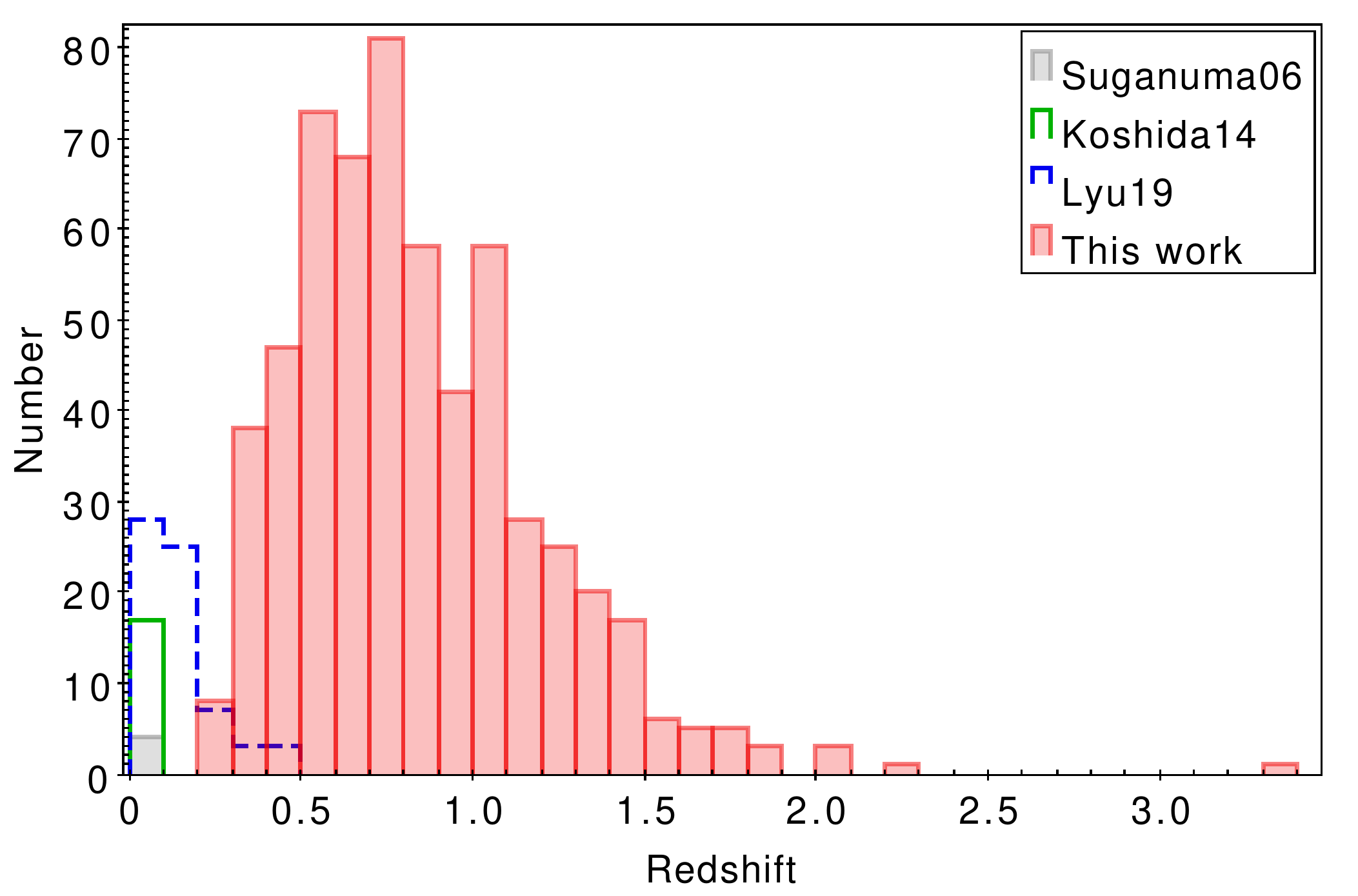}
\caption{Redshift distributions of different AGN/quasar samples with IR lag measurements. \label{fig:redshift}}
\end{figure}

Figure~\ref{fig:redshift} displays the redshift distribution for various samples with IR lag measurements. Our sample covers a much broader redshift range and is the first statistical sample of IR lags at $z\gtrsim 0.3$. 

\section{Results} \label{sec:results}

\begin{figure}
\centering
\includegraphics[width=0.48\textwidth]{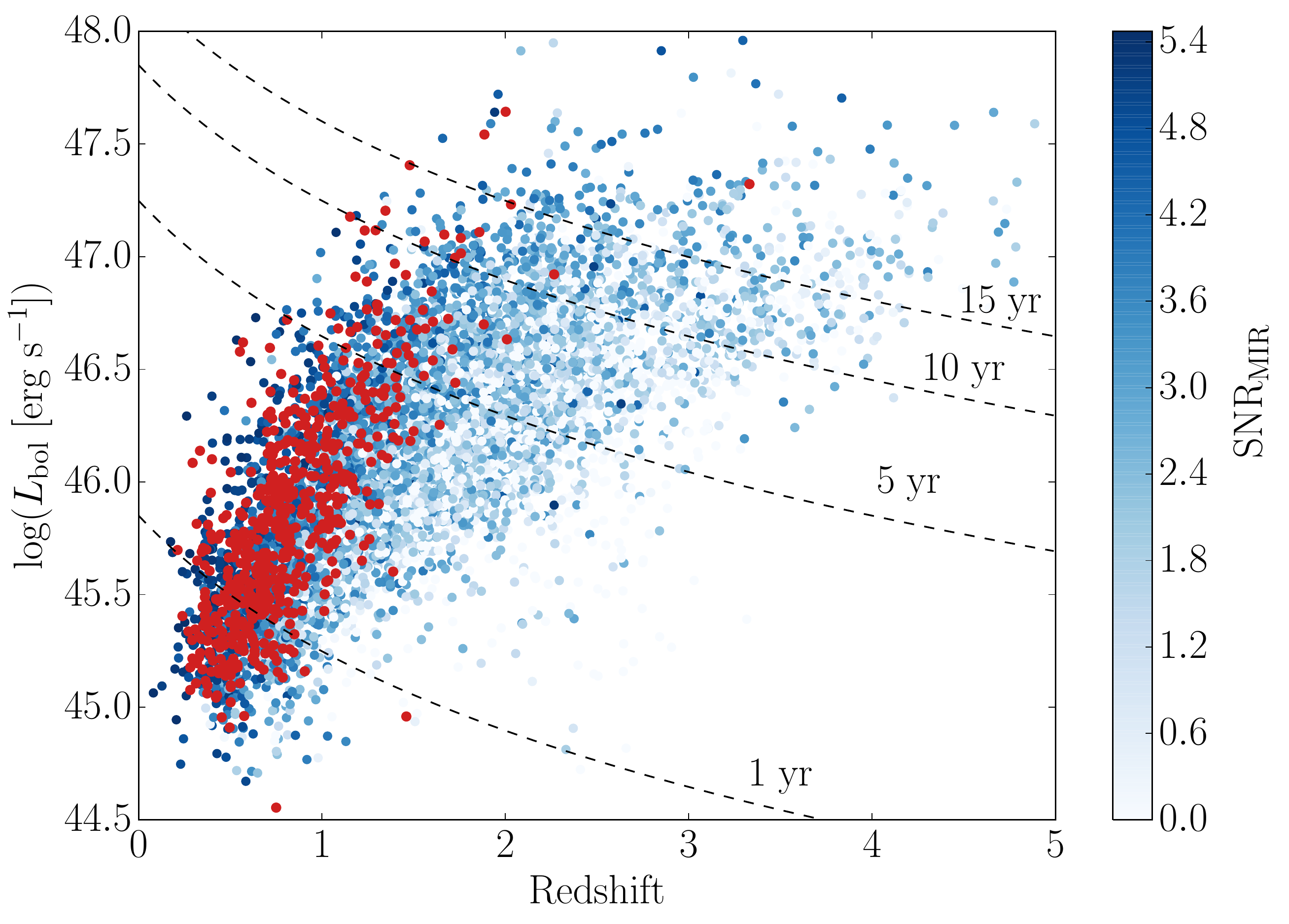}
\caption{Bolometric luminosity versus redshift for our parent quasar sample (blue points) and the final lag sample (red points). The four dashed lines correspond to observed-frame lags of 1 yr, 5 yr, 10 yr and 15 yr, respectively, using the best-fit $R-L$ relation in \citet{Kishimoto2007}. The blue dots are color-coded by the W1 intrinsic variability ${\rm SNR_{MIR}}$. Our final lag sample only includes those with ${\rm SNR_{MIR}}>4$ and have lags that are measurable with the baselines of our light curves. The few objects scattered beyond the $\tau_{\rm obs}=15\,$yr line have shorter actual lags ($<15$\,yr) than predicted from the canonical $R-L$ relation in \citet{Kishimoto2007}. \label{fig:Lz}} 
\end{figure}

\begin{figure}
\centering
\includegraphics[width=0.48\textwidth]{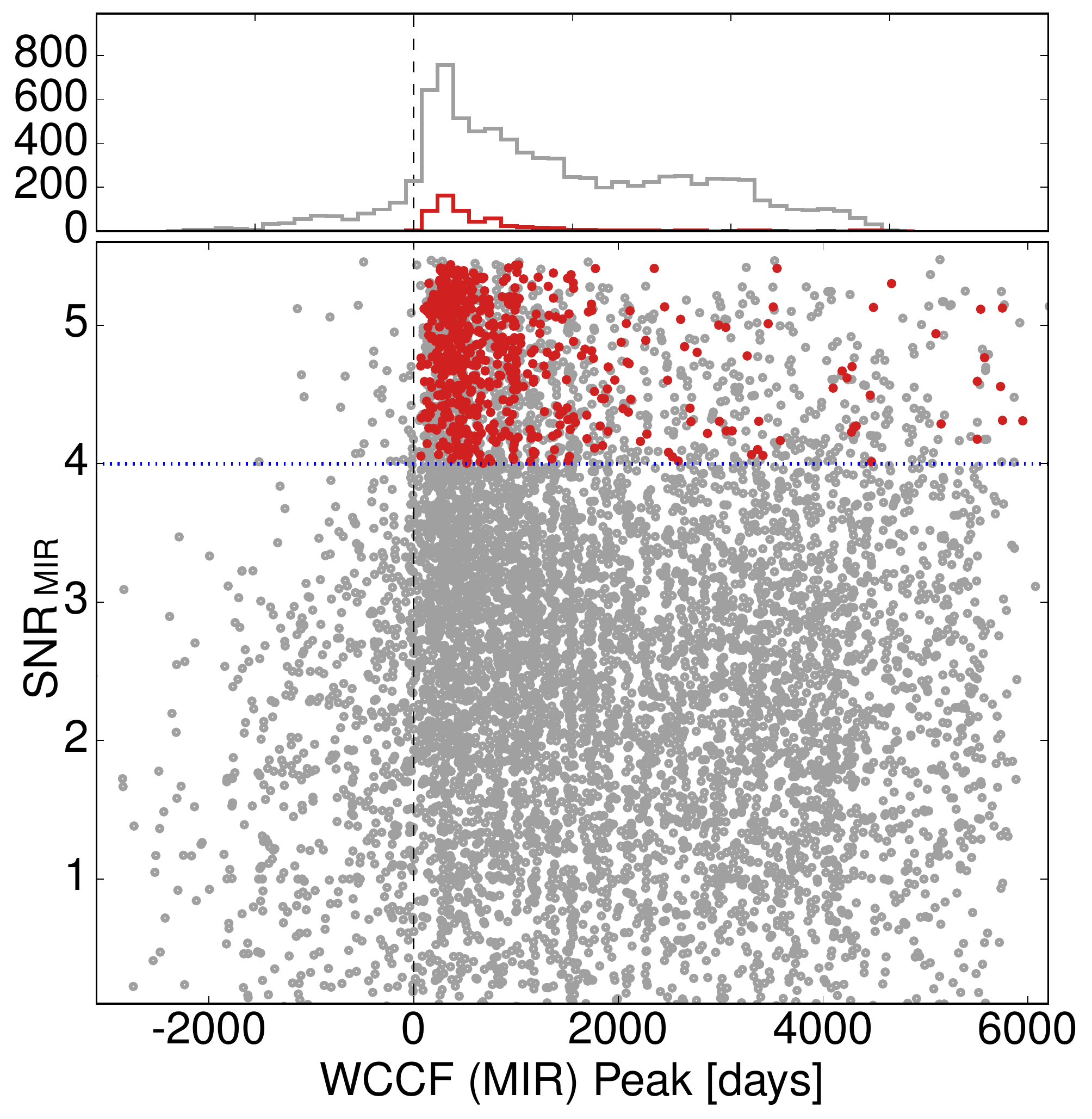}
\caption{${\rm SNR_{MIR}}$ of W1 intrinsic rms versus the peak of WCCF for the parent sample (gray points) and the final lag sample (red points). The top panel shows the histograms of both samples. The asymmetry below the SNR$=4$ cut suggests that most of these lags are genuine lags in a statistical sense, albeit with larger uncertainties compared with the high-fidelity lags shown in red. \label{fig:wccf_snr}}
\end{figure}

We show our main results using the finalcut sample of 587 quasars with high-fidelity MIR lags in Fig.~\ref{fig:RL}. We have converted all luminosities to V-band luminosity (assuming a bolometric correction of 10 in V-band) to directly compare with earlier studies on low-redshift AGN and PG quasars. Our sample well samples the high-luminosity end of the distribution, and tightly follow the best-fit relation in \cite{Koshida2014} based on a local AGN sample and K-band lags. Our MIR lags also agree with the lags measured for tens of PG quasars in \citet{Lyu2019} over the same luminosity range, indicating negligible evolution from $z<0.5$ to $z\sim 1$. 

Because our sample dominates in number in the $R-L$ plot, a joint fit of all lags across the entire luminosity range will be heavily weighted by our sample. Furthermore, several selection effects will tend to bias the measurement of the $R-L$ relation, in particular the slope (see discussions in \S\ref{sec:disc}). For these reasons, we do not perform a formal fit to the IR $R-L$ relation in this work, and defer a proper analysis of the $R-L$ relation to future work. Nevertheless, we can still estimate the intrinsic scatter of this relation using our large sample, which is much less biased than the slope due to selection effects. Using the Bayesian regression algorithm developed by \citet{Kelly_2007}, we estimate an intrinsic scatter of only $\sim 0.17$\,dex for our IR lag sample.

Fig.~\ref{fig:RL} demonstrates the power of combining decade-long optical and IR surveys in dust reverberation mapping in distant quasars. The inferred $R-L$ relation over more than four orders of magnitude in AGN luminosity has profound implications for torus structure and physics, as well as the utility of using this relation as a luminosity indicator for cosmology. We discuss potential selection biases in our sample and origins of the scatter in the relation in \S\ref{sec:disc}.

\section{Discussion}\label{sec:disc}

\begin{figure*}
\centering
\includegraphics[width=0.8\textwidth]{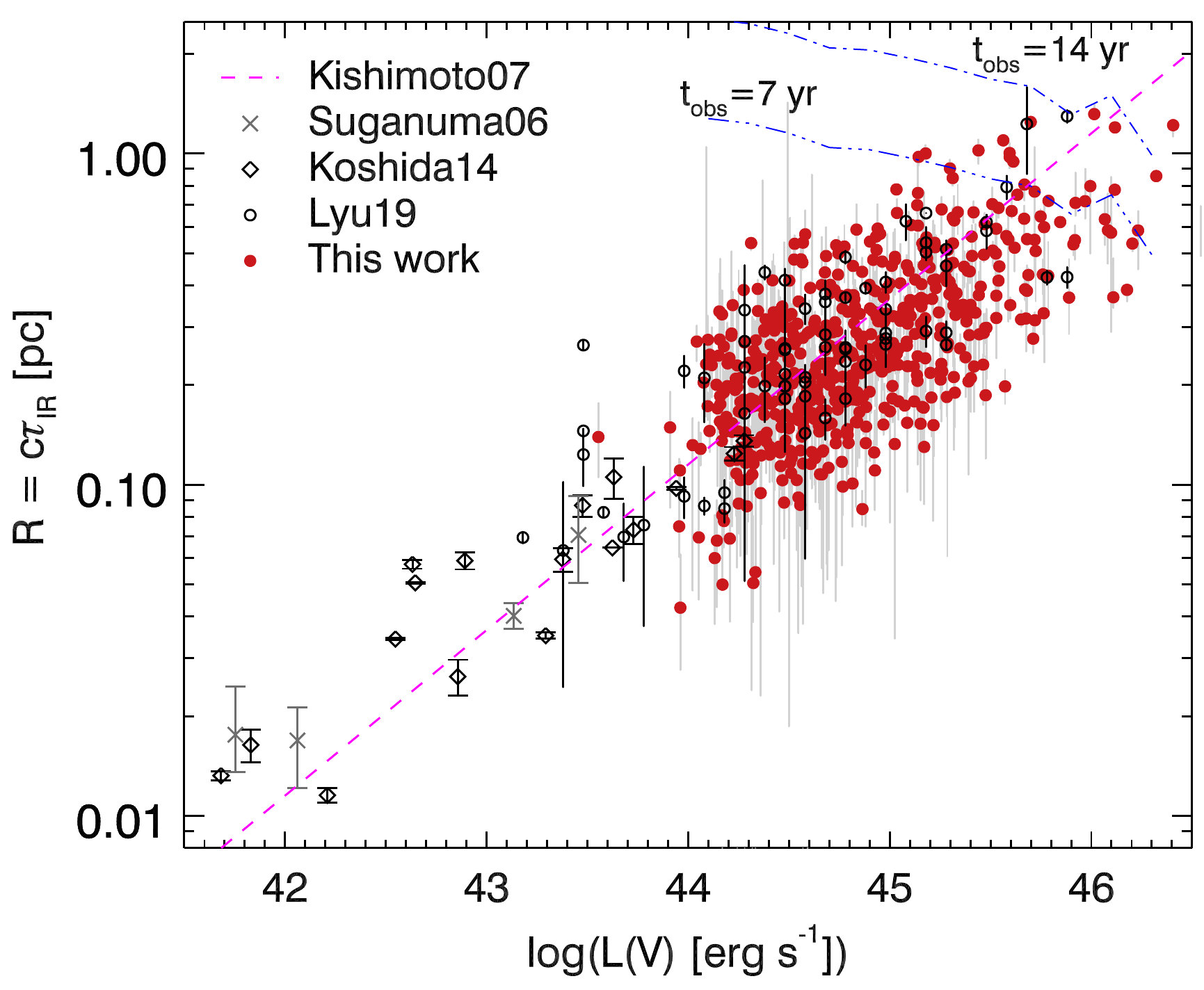}
\caption{The correlation between the torus size (inferred from the rest-frame lag $\tau_{\rm IR}$) and the rest-frame optical quasar luminosity. Our finalcut lag sample is shown in red filled circles with gray error bars, and the PG quasars from \citet{Lyu2019} are shown in open circles. The black diamonds and gray crosses are earlier lag measurements between $K-$band and optical band for a small sample of local AGN \citep{Suganuma2006,Koshida2014}. The magenta line is the $R-L$ relation in \citet{Kishimoto2007}. The two blue dash-dotted lines are the rest-frame timescale corresponding to the marked observed-frame time duration (7 yr and 14 yr) calculated at the 90th percentile redshift of our quasars in each luminosity bin. Regions above the $t_{\rm obs}=7\,{\rm yr}$ line will suffer significant incompleteness in lag detection given the lengths of our optical and MIR light curves.  \label{fig:RL}}
\end{figure*}

Fig.~\ref{fig:RL} reveals a slight trend of deviation from the local $R-L$ relation towards the high-luminosity end. This deviation is most likely caused by selection effects. First, higher-luminosity quasars on average are at higher redshifts (Fig.~\ref{fig:Lz}), where the longer observed-frame lags due to higher luminosity and the $1+z$ cosmic time dilation are more difficult to measure given the fixed light curve baseline. Modeling the detailed selection function given the light curve duration, cadence and variability S/N and for an underlying sample of quasars over broad redshift and luminosity ranges can be achieved with simulations \citep[e.g.,][]{Shen_etal_2015a,Li_etal_2019}, but is beyond the scope of this paper. Here we provide a qualitative understanding of the loss of long lags due to the baselines of our optical+MIR light curves.

In a given luminosity bin for our quasar sample in Fig.~\ref{fig:RL}, we determine the 90th percentile redshift, $z_{90}$, in that bin, and calculate the rest-frame time at $z_{90}$ corresponding to a given observed-frame time duration. We use this rest-frame time as a rough estimate of the upper limit of measurable lags at that luminosity, given the survey length. Since higher-luminosity quasars on average have higher redshifts in our sample, this rest-frame timescale decreases with luminosity. Given the actual lengths of our optical and MIR light curves, we can reasonably assume that all lags shorter than 7 yrs in the observed frame will be detected since the shifted MIR light curve is in complete overlap with the earlier optical light curve. However, the probability of detection will decrease towards longer lags in the observed frame. For an observed-frame lag of 14 yrs, we only have $\sim60\%$ of the full MIR light curve overlapping with the earlier optical light curve and therefore the detection probability will be significantly reduced compared to the case of shorter lags. Therefore we expect our survey will start significantly losing lags beyond the line corresponding to 7 yrs, shown in Fig.~\ref{fig:RL} as the dash-dot-dot-dotted line. As a consequence, this selection effect will bias the average lags low towards the high-luminosity end. 

A second selection effect is that as objects move towards the high-luminosity end (on average high-redshift end), the W1 3.5\,\micron\ band samples shorter rest-frame IR wavelength. We have found that (Yang et~al., in prep) the W1-band lags are systematically shorter than the W2-band lags in the same quasars, suggesting wavelength-dependent IR lags. This selection effect would bias the average lags lower towards higher redshift and therefore higher average luminosity in Fig.~\ref{fig:RL}, leading to slight flattening of the slope. Non-negligible contamination from the accretion disk in the W1 band in extreme cases may also contribute to this bias. 

The combination of these two selection effects can qualitatively explain the slight deviation of our sample from the local $R-L$ relation at the high-luminosity end. In future work, we will carefully model these selection effects with detailed simulations, and to derive unbiased constraints on the IR $R-L$ relation.  

\begin{figure*}
\centering
  \includegraphics[width=0.48\textwidth]{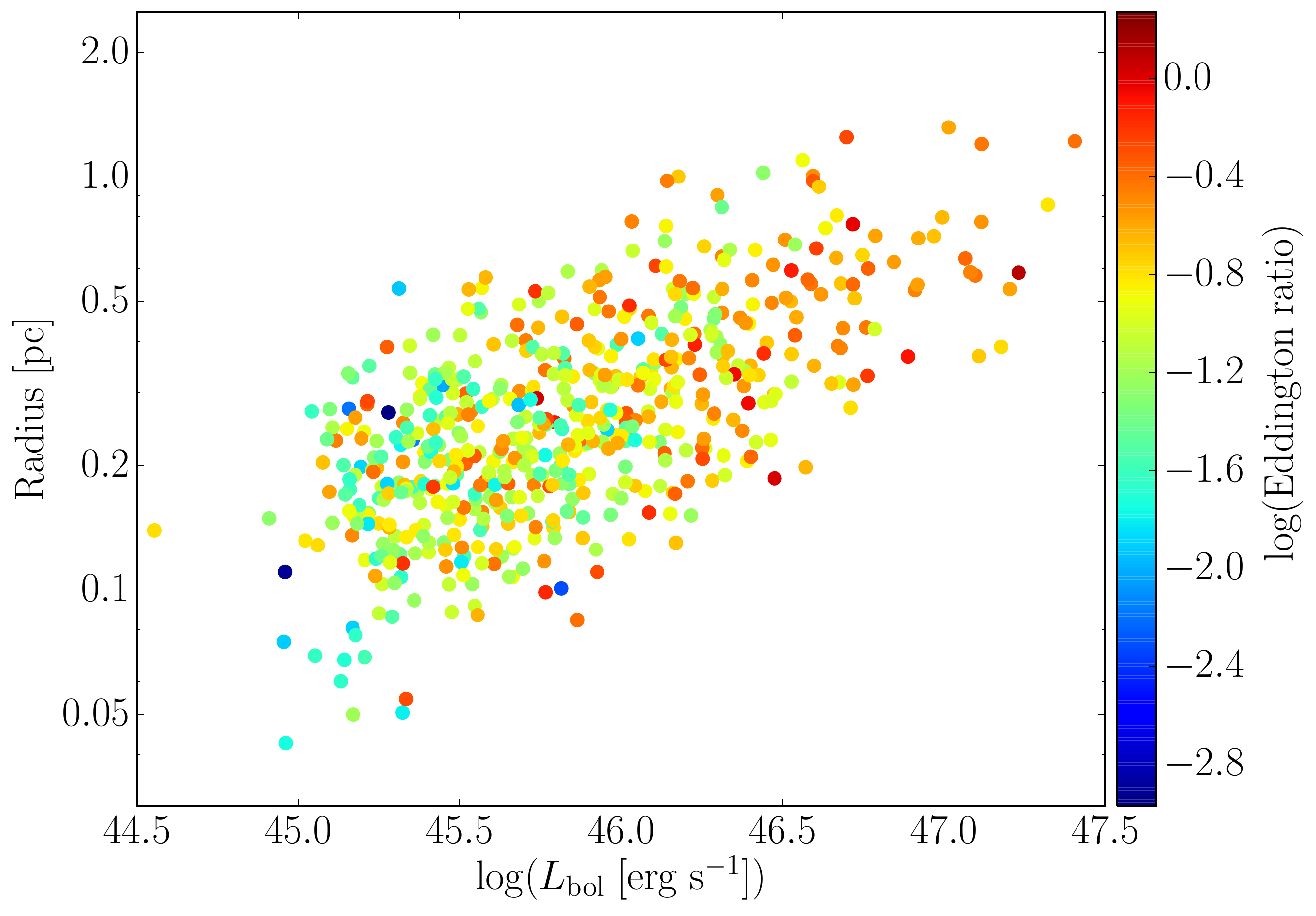}
  \includegraphics[width=0.48\textwidth]{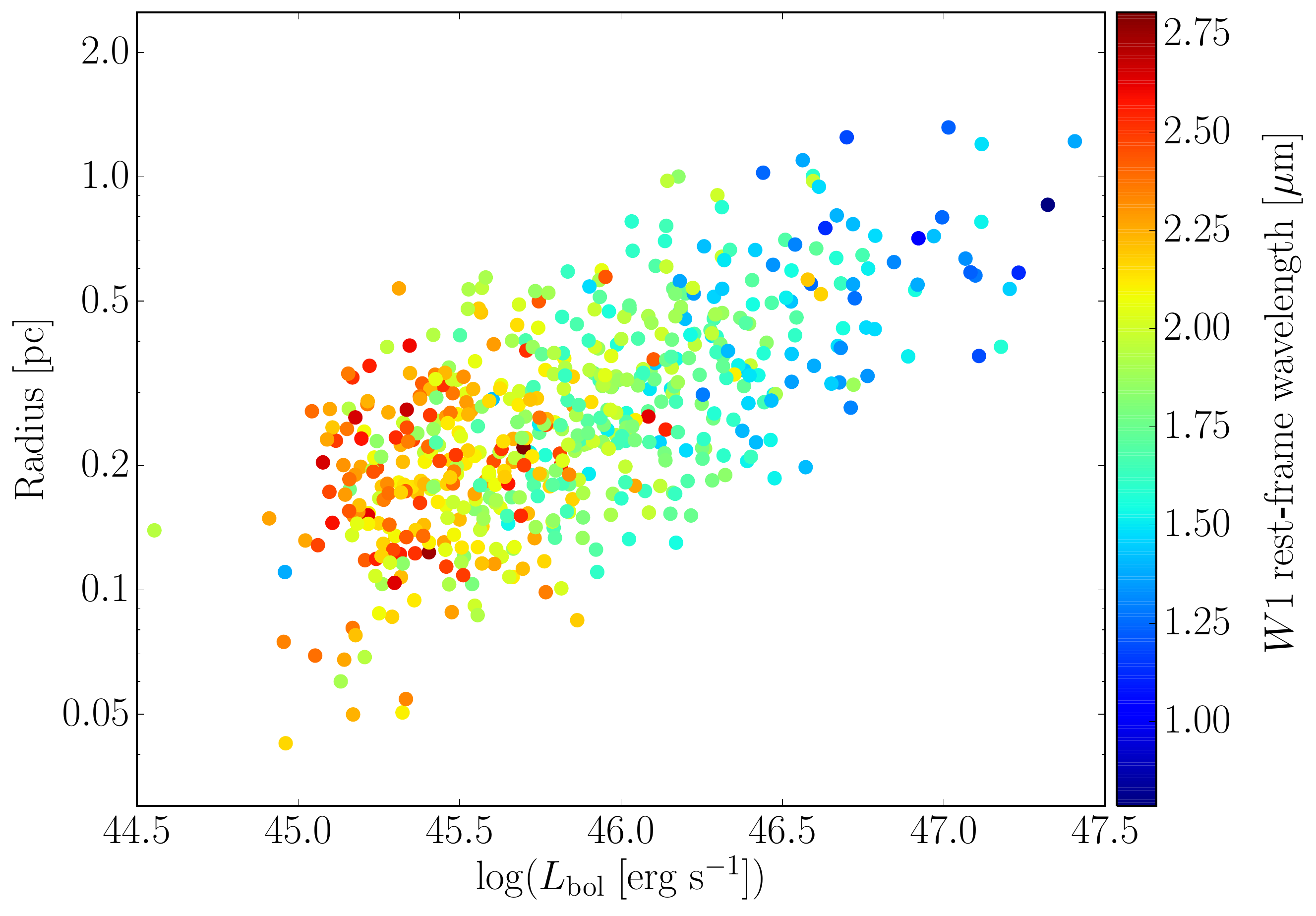}
  \includegraphics[width=0.48\textwidth]{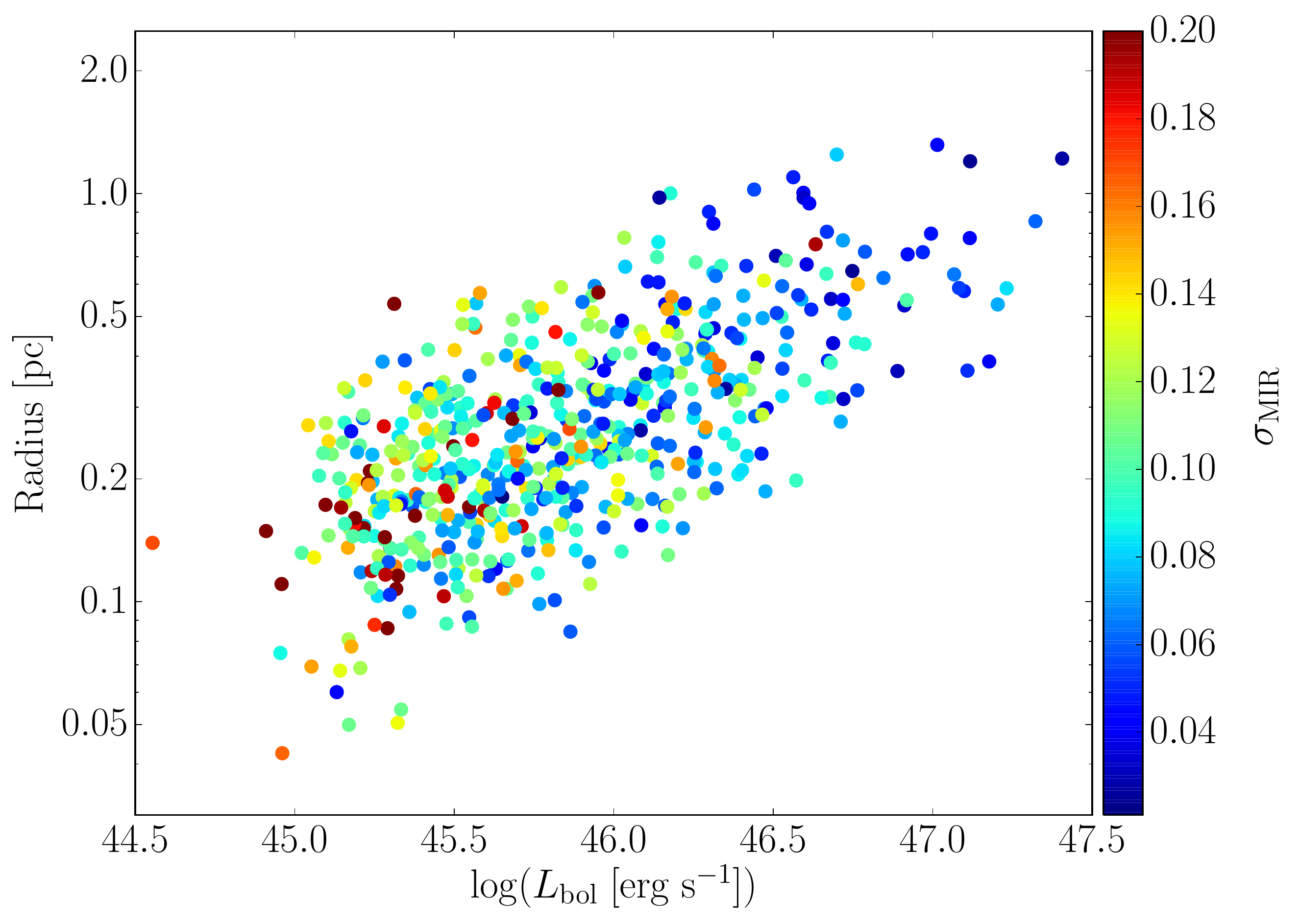}
  \caption{Distributions of our finalcut lag sample in the $R-L$ plane, color-coded by different properties (top-left: Eddington ratio; top-right: rest-frame wavelength sampled by W1; bottom: intrinsic W1 variability in magnitude). At fixed luminosity, the scatter in lag does not depend on any of these additional parameters. The apparent trend of rest-frame wavelength with luminosity is driven by the apparent redshift-luminosity relation. The apparent trend of reducing average variability amplitude with luminosity is consistent with the variability-luminosity relation observed in the optical.  \label{fig:RL_version1}}
\end{figure*}

To examine the scatter around the average $R-L$ relation, we show our finalcut lag sample in Fig.~\ref{fig:RL_version1}, where we color-code the objects by various properties. We found that the scatter in MIR lag at fixed luminosity does not depend on the Eddington ratio or the variability amplitude of the quasar. This greatly simplifies the interpretation of the MIR $R-L$ relation as due to a primary luminosity-driven effect. The apparent trend of rest-frame IR wavelength sampled by W1 as a function of luminosity (upper-right panel of Fig.~\ref{fig:RL_version1}) is caused by the apparent luminosity-redshift relation. In addition, higher-luminosity quasars tend to have on average lower variability amplitude (bottom panel), consistent with previous studies of quasar variability in MIR \citep[e.g.,][]{Kozlowski2016} and optical \citep[e.g.,][]{MacLeod_etal_2012}.

\section{Summary} \label{sec:summary}

We have presented first results from our dust reverberation mapping study of distant ($z\gtrsim 0.3$) quasars in the SDSS Stripe 82 region, using ground-based optical imaging and the MIR light curves from the WISE satellite. Our optical light curves span a baseline of $\sim 20$ years and the MIR light curves span almost a decade. We presented high-fidelity optical-WISE W1 (3.5\,\micron) lag measurements for 587 quasars at $0.3\lesssim z\lesssim 2$, with a median redshift of $0.8$. Our statistical analysis suggests there are thousands more lags that are measurable (Fig.~\ref{fig:wccf_snr}), but their uncertainties are much larger than our high-fidelity sample, and therefore these less reliable lags are excluded from our analysis in this work. 

The 587 quasars with high-fidelity MIR lags span more than two orders of magnitude in quasar luminosity, and tightly follow the torus $R-L$ relation observed for nearby AGN observed in the optical and K-band. The intrinsic scatter of the $R-L$ relation defined by our sample alone is only $\sim 0.17$\,dex. Furthermore, there is no apparent dependence of the scatter in IR lag at fixed luminosity on additional quasar properties such as the Eddington ratio and variability amplitude of the quasar, suggesting luminosity is the primary driver to determine the IR lag. However, to robustly measure the $R-L$ relation with the WISE sample, we must account for selection biases at the high luminosity end due to the duration of the light curves (\S\ref{sec:disc}). In addition, the WISE W1 band probes shorter rest-frame wavelengths towards higher redshift, and the wavelength-dependence of torus lags must also be taken into consideration.

The observed global torus $R-L$ relation over more than four orders of magnitude in AGN luminosity is remarkable, considering the small intrinsic scatter and the lack of dependence on additional quasar properties. The physics of setting the inner dust torus radius (e.g., dust sublimation) is also much simpler than photoionization in broad-line RM \citep[e.g.,][]{Czerny_Hryniewicz_2011}, making this dust $R-L$ relation an attractive luminosity indicator to probe cosmology at high redshift \citep[e.g.,][]{Yoshii_etal_2014,Koshida_etal_2017}. However, the systematics in the dust $R-L$ relation and selection biases affecting the measurement of this relation must be thoroughly investigated before this technique can be applied to cosmology.   

On the other hand, measuring reliable dust torus lags in AGN with different physical properties across a broad range of redshift also has tremendous value to understanding the physics of AGN. Our results based on current optial and IR imaging survey data provide strong endorsement for the joint analysis of the 10-yr optical light curves from the Vera C. Rubin Observatory Legacy Survey of Space and Time (LSST) and the planned 5-yr IR survey with the Nancy Grace Roman Space Telescope (NGRST, previously known as WFIRST). The LSST+NGRST data set resembles the light curves studied here with a similar shift in the onset of the optical and IR temporal coverage. However, LSST+NGRST will be able to measure light curves for AGN that are up to 5 magnitudes fainter than our S82 quasar sample. The expected dust lags from these lower-luminosity AGN will be $\sim$a factor of ten shorter, and therefore easily measurable with the maximum $\sim 10$-yr LSST+NGRST baselines. Since NGRST only goes to $\sim 2$\,\micron, the dust torus lag measurements are necessarily limited to the $z\lesssim 0.5$ regime, where the variability at $2$\,\micron\ (observed-frame) from the torus emission can still be reasonably well measured on top of the constant host galaxy light. Importantly, by optimizing the cadence and overlap between NGRST and LSST, it is possible to compile a large sample of low-redshift and low-luminosity AGN with well measured torus lags to densely populate the low-luminosity regime in the torus $R-L$ relation. For reference, the cumulative sky density of $i<24$ quasars at $z<0.5$ is $\sim 30\,{\rm deg^{-2}}$ estimated using the extrapolation of the \citet{Hopkins_etal_2007} quasar luminosity function. The V-band luminosity for $z=0.5$ AGN at this flux limit is about $\log (L_V/{\rm erg\,s^{-1}})\sim 42.4$ using magnitude and luminosity conversions in \citet[][]{Richards_etal_2006a} and \citet{Shen_etal_2009}, corresponding to an observed-frame torus lag of $\sim 1$ month. Thus even the wide survey cadence of LSST \citep[e.g., $15$\,days in $r$ and 3\,days in merged multi-band light curves,][]{LSSTObservingStrategyWhitePaper} would be sufficient for lag measurements. The LSST Deep Drill Fields will have much higher cadences over tens of square degrees, which could be the high-priority fields for NGRST to (partially) overlap in a medium-deep survey and to provide dense IR light curves.

We outline some ongoing work to further explore the utility of current optical+WISE light curves on AGN dust reverberation mapping: 

\begin{enumerate}
    \item[$\bullet$] We are developing more robust forward modeling scheme to measure the intrinsic torus $R-L$ relation from surveys that takes into account selection effects introduced by the duration (and to a lesser extent, cadence) of light curves, using simulated light curves following the approaches described in e.g., \citet[][]{Shen_etal_2015a} and \citet{Li_etal_2019}. 
    
    \item[$\bullet$] We are studying individual cases with sufficient light curve quality to constrain the transfer function of dust reverberation mapping, and developing physical models for the structure of AGN dust torus that can be constrained with dust RM \citep[e.g.,][]{Nenkova2008,Shen_2012,Almeyda_etal_2017,Yang_etal_2019b}.  
    
    \item[$\bullet$] We are investigating multi-band IR lags (W1 and W2) for the S82 sample, and will use these multi-band lags to further constrain the temperature profile of the torus \citep[e.g.,][]{Honig_Kishimoto_2011} and the induced scatter in the $R-L$ relation. In the meantime, we are compiling NIR light curves for a subset of our quasars to measure NIR lags to extend the wavelength coverage. 
    
    \item[$\bullet$] We are expanding our dust RM analysis to quasars outside the Stripe 82 region. The longest optical baseline then comes from the shallower CRTS survey, which necessarily limits our study to the rarer and brighter quasars. However, the much larger sky coverage compensates for the lower sky density and therefore we expect a significant increase in the number of the brightest quasars with measurable lags from optical and WISE light curves, most of which will be at lower redshifts and lower luminosities than the S82 sample studied here.  
    
\end{enumerate}

\clearpage
\acknowledgments

We thank the anonymous referee for comments that improved the manuscript. QY and YS acknowledge support from NSF grant AST-1715579 (QY, YS) and an Alfred P. Sloan Research Fellowship (YS). 

This publication makes use of data products from the {\it WISE}, which is a joint project of the University of California, Los Angeles, and the Jet Propulsion Laboratory/California Institute of Technology, funded by the National Aeronautics and Space Administration. This publication also makes use of data products from {\it NEOWISE}, which is a project of the Jet Propulsion Laboratory/California Institute of Technology, funded by the Planetary Science Division of the National Aeronautics and Space Administration.

Funding for the DES Projects has been provided by the U.S. Department of Energy, the U.S. National Science Foundation, the Ministry of Science and Education of Spain, the Science and Technology Facilities Council of the United Kingdom, the Higher Education Funding Council for England, the National Center for Supercomputing Applications at the University of Illinois at Urbana-Champaign, the Kavli Institute of Cosmological Physics at the University of Chicago, the Center for Cosmology and Astro-Particle Physics at the Ohio State University, the Mitchell Institute for Fundamental Physics and Astronomy at Texas A\&M University, Financiadora de Estudos e Projetos, Funda{\c c}{\~a}o Carlos Chagas Filho de Amparo {\`a} Pesquisa do Estado do Rio de Janeiro, Conselho Nacional de Desenvolvimento Cient{\'i}fico e Tecnol{\'o}gico and 
the Minist{\'e}rio da Ci{\^e}ncia, Tecnologia e Inova{\c c}{\~a}o, the Deutsche Forschungsgemeinschaft and the Collaborating Institutions in the Dark Energy Survey. 

The Collaborating Institutions are Argonne National Laboratory, the University of California at Santa Cruz, the University of Cambridge, Centro de Investigaciones Energ{\'e}ticas, Medioambientales y Tecnol{\'o}gicas-Madrid, the University of Chicago, University College London, the DES-Brazil Consortium, the University of Edinburgh, the Eidgen{\"o}ssische Technische Hochschule (ETH) Z{\"u}rich, Fermi National Accelerator Laboratory, the University of Illinois at Urbana-Champaign, the Institut de Ci{\`e}ncies de l'Espai (IEEC/CSIC), the Institut de F{\'i}sica d'Altes Energies, Lawrence Berkeley National Laboratory, the Ludwig-Maximilians Universit{\"a}t M{\"u}nchen and the associated Excellence Cluster Universe, 
the University of Michigan, the National Optical Astronomy Observatory, the University of Nottingham, The Ohio State University, the University of Pennsylvania, the University of Portsmouth, SLAC National Accelerator Laboratory, Stanford University, the University of Sussex, Texas A\&M University, and the OzDES Membership Consortium.

Based in part on observations at Cerro Tololo Inter-American Observatory, National Optical Astronomy Observatory, which is operated by the Association of Universities for Research in Astronomy (AURA) under a cooperative agreement with the National Science Foundation.

Funding for the SDSS and SDSS-II has been provided by the Alfred P. Sloan Foundation, the Participating Institutions, the National Science Foundation, the U.S. Department of Energy, the National Aeronautics and Space Administration, the Japanese Monbukagakusho, the Max Planck Society, and the Higher Education Funding Council for England. The SDSS Web Site is http://www.sdss.org/. The SDSS is managed by the Astrophysical Research Consortium for the Participating Institutions. The Participating Institutions are the American Museum of Natural History, Astrophysical Institute Potsdam, University of Basel, University of Cambridge, Case Western Reserve University, University of Chicago, Drexel University, Fermilab, the Institute for Advanced Study, the Japan Participation Group, Johns Hopkins University, the Joint Institute for Nuclear Astrophysics, the Kavli Institute for Particle Astrophysics and Cosmology, the Korean Scientist Group, the Chinese Academy of Sciences (LAMOST), Los Alamos National Laboratory, the Max-Planck-Institute for Astronomy (MPIA), the Max-Planck-Institute for Astrophysics (MPA), New Mexico State University, Ohio State University, University of Pittsburgh, University of Portsmouth, Princeton University, the United States Naval Observatory, and the University of Washington.

The PS1 and the PS1 public science archive have been made possible through contributions by the Institute for Astronomy, the University of Hawaii, the Pan-STARRS Project Office, the Max-Planck Society and its participating institutes, the Max Planck Institute for Astronomy, Heidelberg and the Max Planck Institute for Extraterrestrial Physics, Garching, The Johns Hopkins University, Durham University, the University of Edinburgh, the Queen's University Belfast, the Harvard-Smithsonian Center for Astrophysics, the Las Cumbres Observatory Global Telescope Network Incorporated, the National Central University of Taiwan, the Space Telescope Science Institute, the National Aeronautics and Space Administration under Grant No. NNX08AR22G issued through the Planetary Science Division of the NASA Science Mission Directorate, the National Science Foundation Grant No. AST-1238877, the University of Maryland, Eotvos Lorand University (ELTE), the Los Alamos National Laboratory, and the Gordon and Betty Moore Foundation.

The CSS survey is funded by the National Aeronautics and Space
Administration under Grant No. NNG05GF22G issued through the Science Mission Directorate Near-Earth Objects Observations Program.  The CRTS survey is supported by the U.S.~National Science Foundation under grants AST-0909182.

ZTF: Based on observations obtained with the Samuel Oschin 48-inch Telescope at the Palomar Observatory as part of the Zwicky Transient Facility project. ZTF is supported by the National Science Foundation under Grant No. AST-1440341 and a collaboration including Caltech, IPAC, the Weizmann Institute for Science, the Oskar Klein Center at Stockholm University, the University of Maryland, the University of Washington, Deutsches Elektronen-Synchrotron and Humboldt University, Los Alamos National Laboratories, the TANGO Consortium of Taiwan, the University of Wisconsin at Milwaukee, and Lawrence Berkeley National Laboratories. Operations are conducted by COO, IPAC, and UW. 

ASAS-SN is supported by the Gordon and Betty Moore Foundation through grant GBMF5490 to the Ohio State University and NSF grant AST-1515927. Development of ASAS-SN has been supported by NSF grant AST-0908816, the Mt. Cuba Astronomical Foundation, the Center for Cosmology and AstroParticle Physics at the Ohio State University, the Chinese Academy of Sciences South America Center for Astronomy (CASSACA), the Villum Foundation, and George Skestos.

\appendix
\setcounter{figure}{0} 
\renewcommand{\thefigure}{A.\arabic{figure}}

\begin{figure}
\centering
\includegraphics[width=0.48\textwidth]{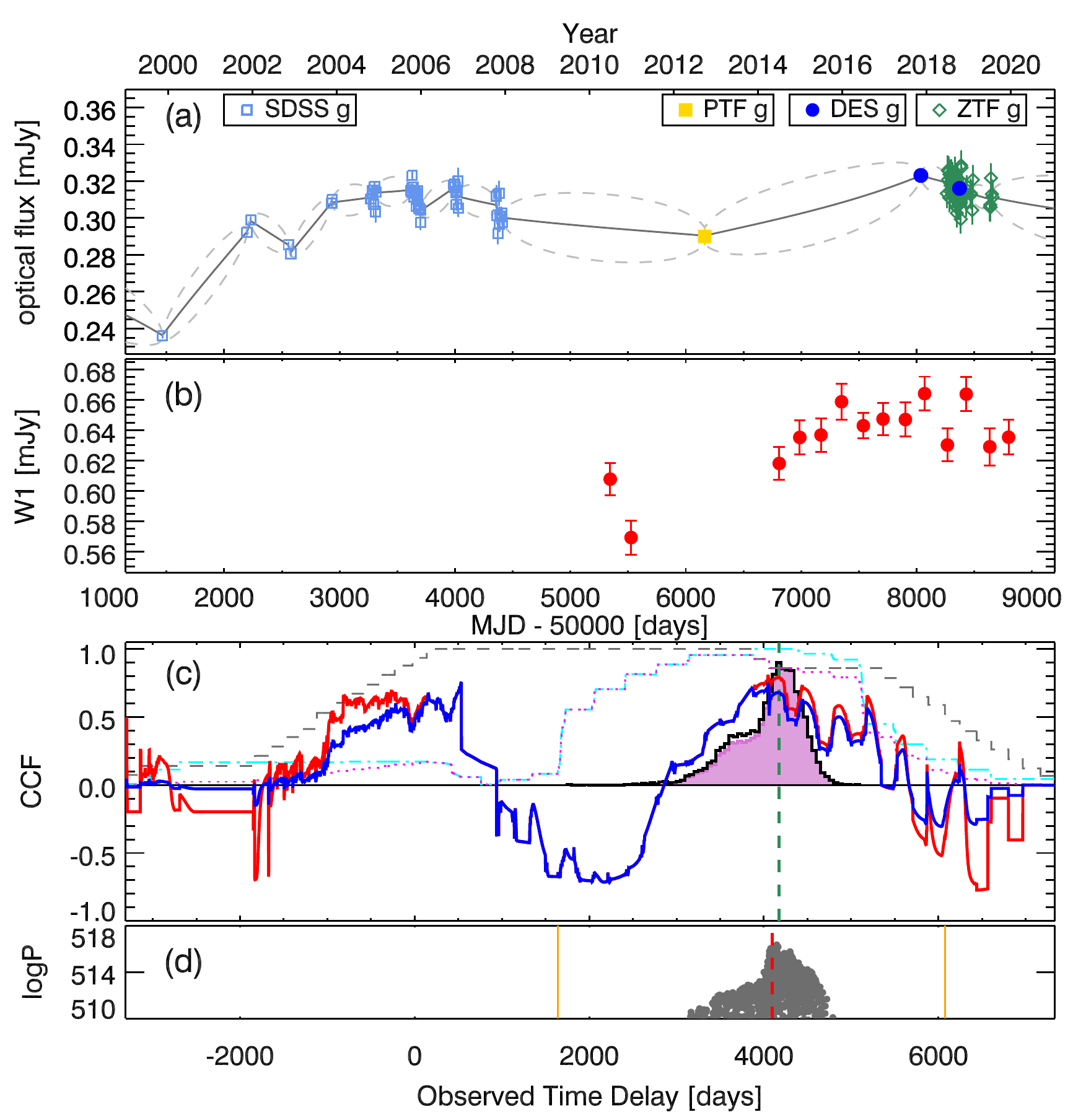}
\includegraphics[width=0.48\textwidth]{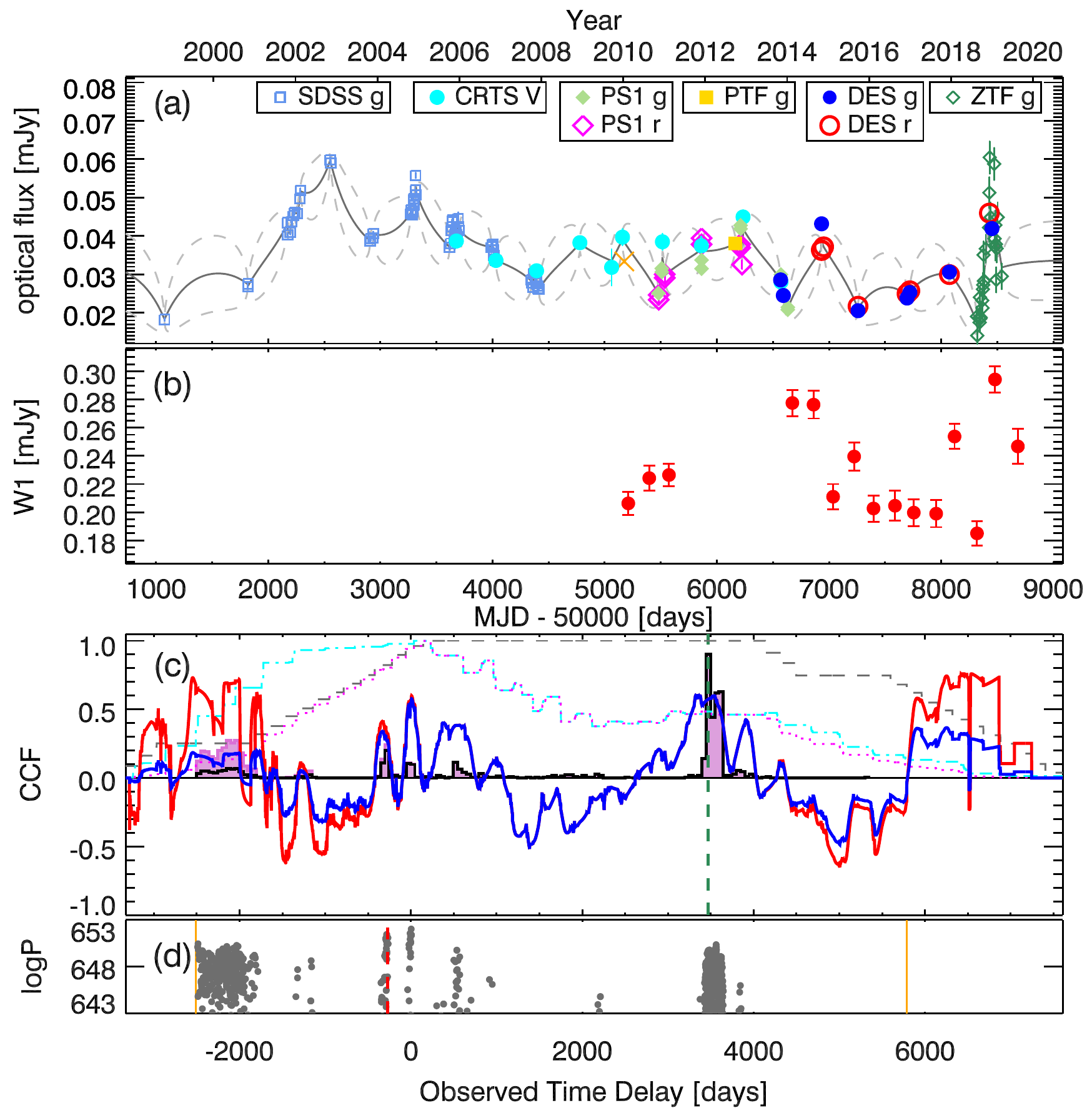}
\includegraphics[width=0.48\textwidth]{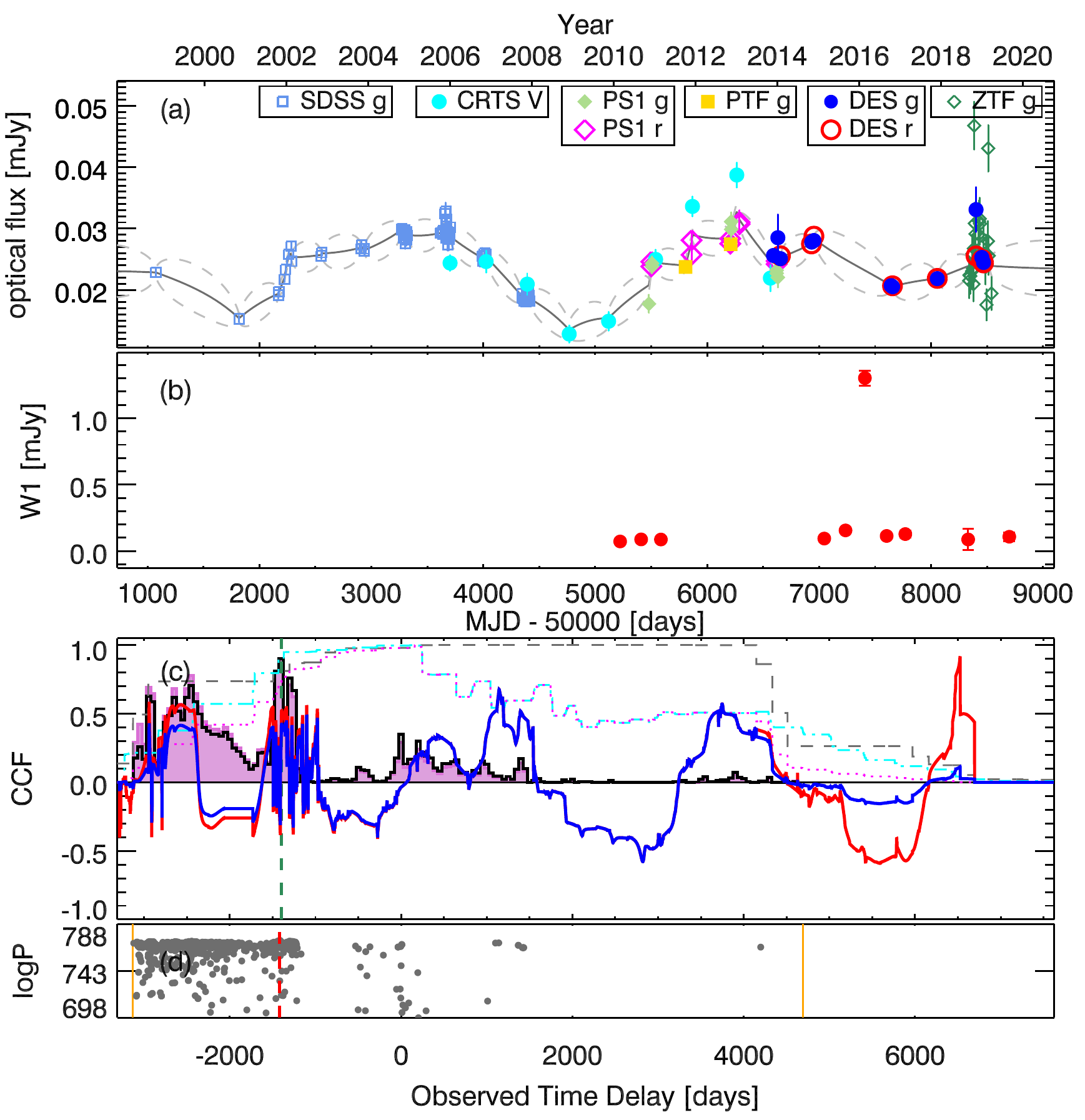}
\includegraphics[width=0.48\textwidth]{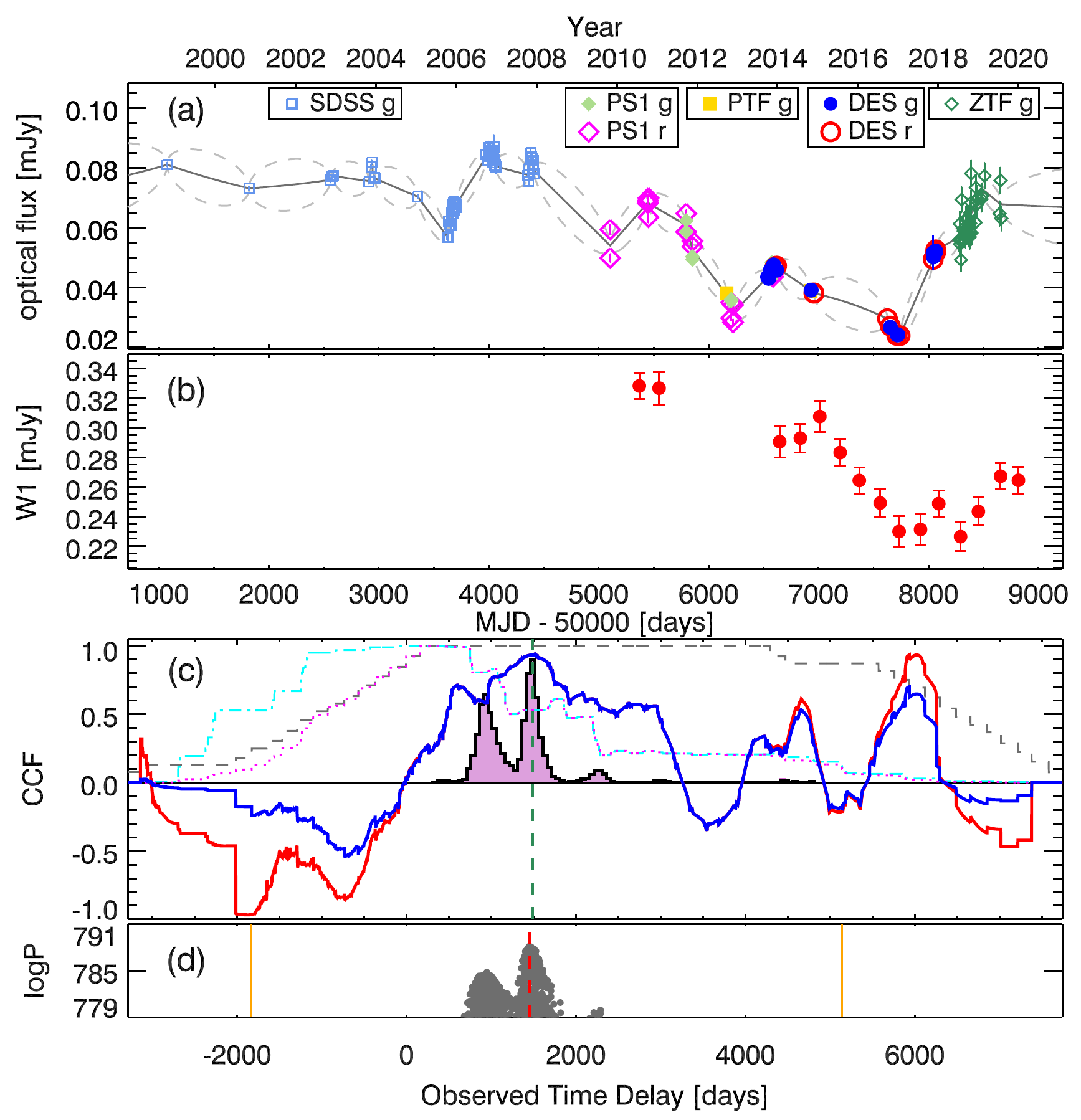}
\caption{{\it Upper-left}: This object was rejected by visual inspection due to large gaps ($>$ 4 yr) in the optical light curve. {\it Upper-right:} This object was rejected by visual inspection because the \javelin\ posterior MCMC lag with the maximum probability (red dash line in panel (d)) is inconsistent with the primary peak in the lag posterior (green dash line in panel (c)). {\it Bottom-left:} This object was rejected by visual inspection because its MIR light curve shows extreme variability in a single epoch and the ICCF is very noisy. {\it Bottom-right:} This object was rejected by visual inspection because there is an obvious secondary peak. This is to ensure that we only select objects with one significant primary peak. 
\label{fig:gap}}
\end{figure}

In Figure \ref{fig:gap} we show one example for each category of objects excluded by visual inspection in our finalcut (\S\ref{sec:finalcut}).

\begin{figure}
\centering
\includegraphics[width=0.48\textwidth]{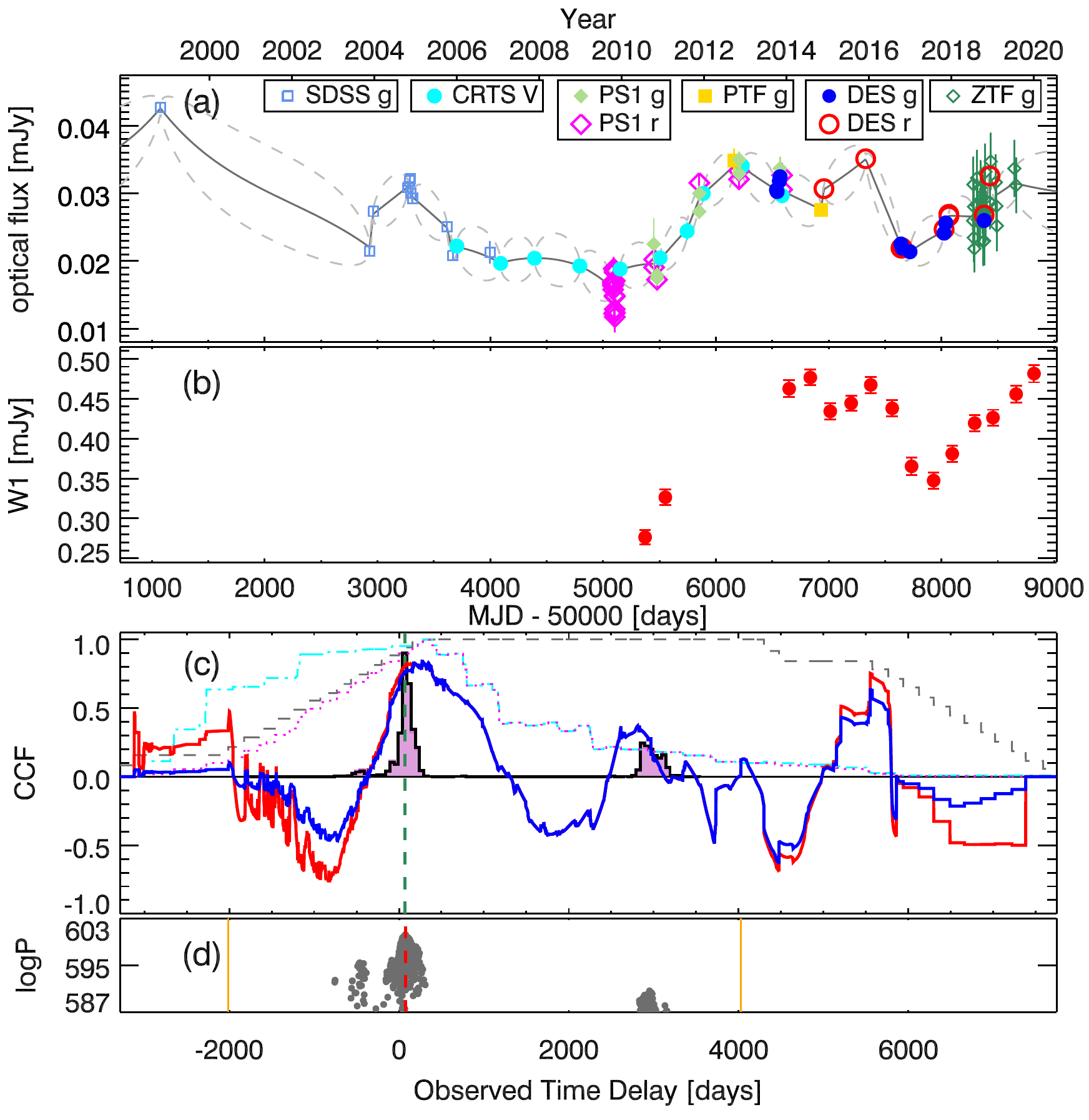}
\includegraphics[width=0.48\textwidth]{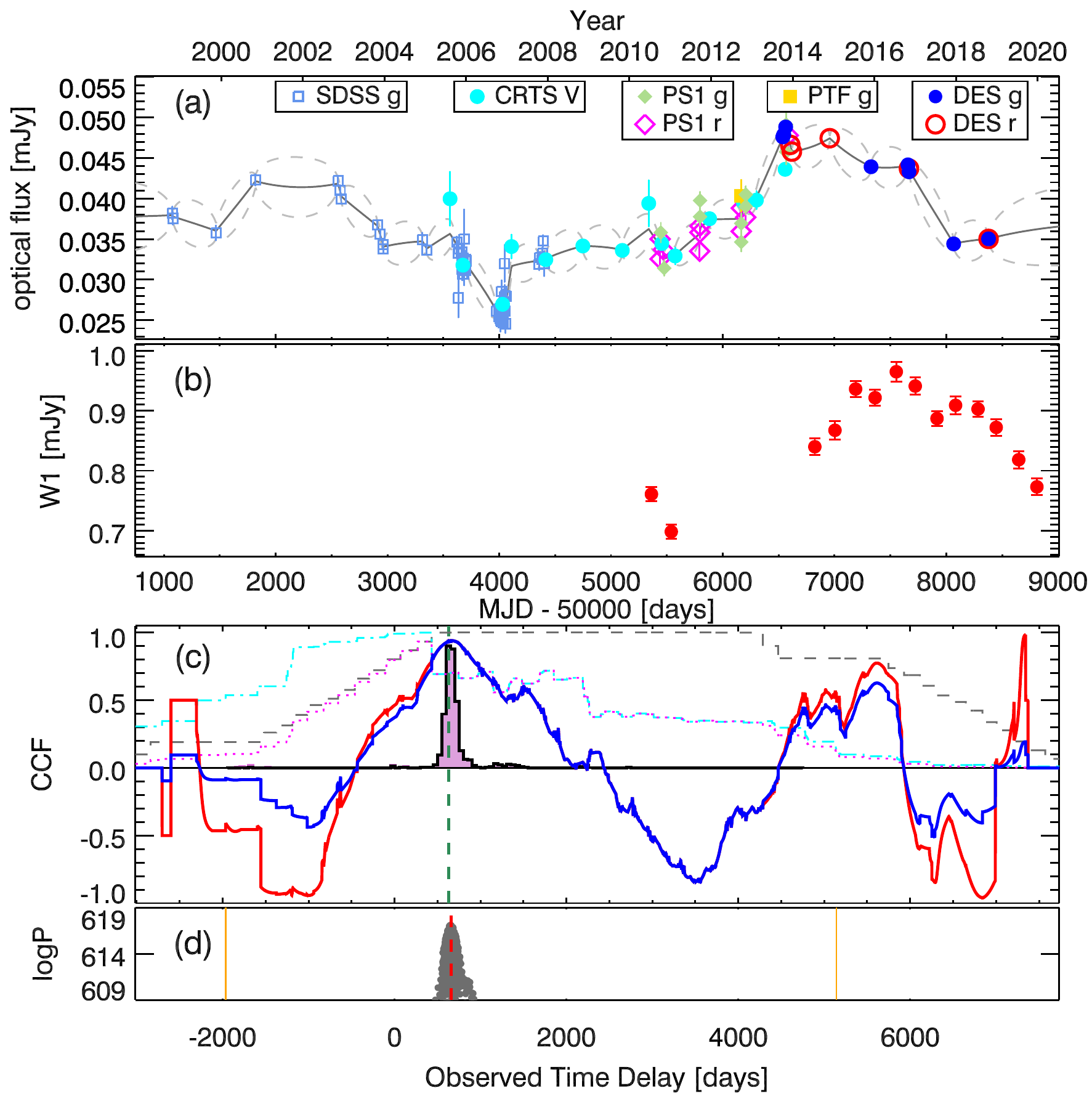}
\includegraphics[width=0.48\textwidth]{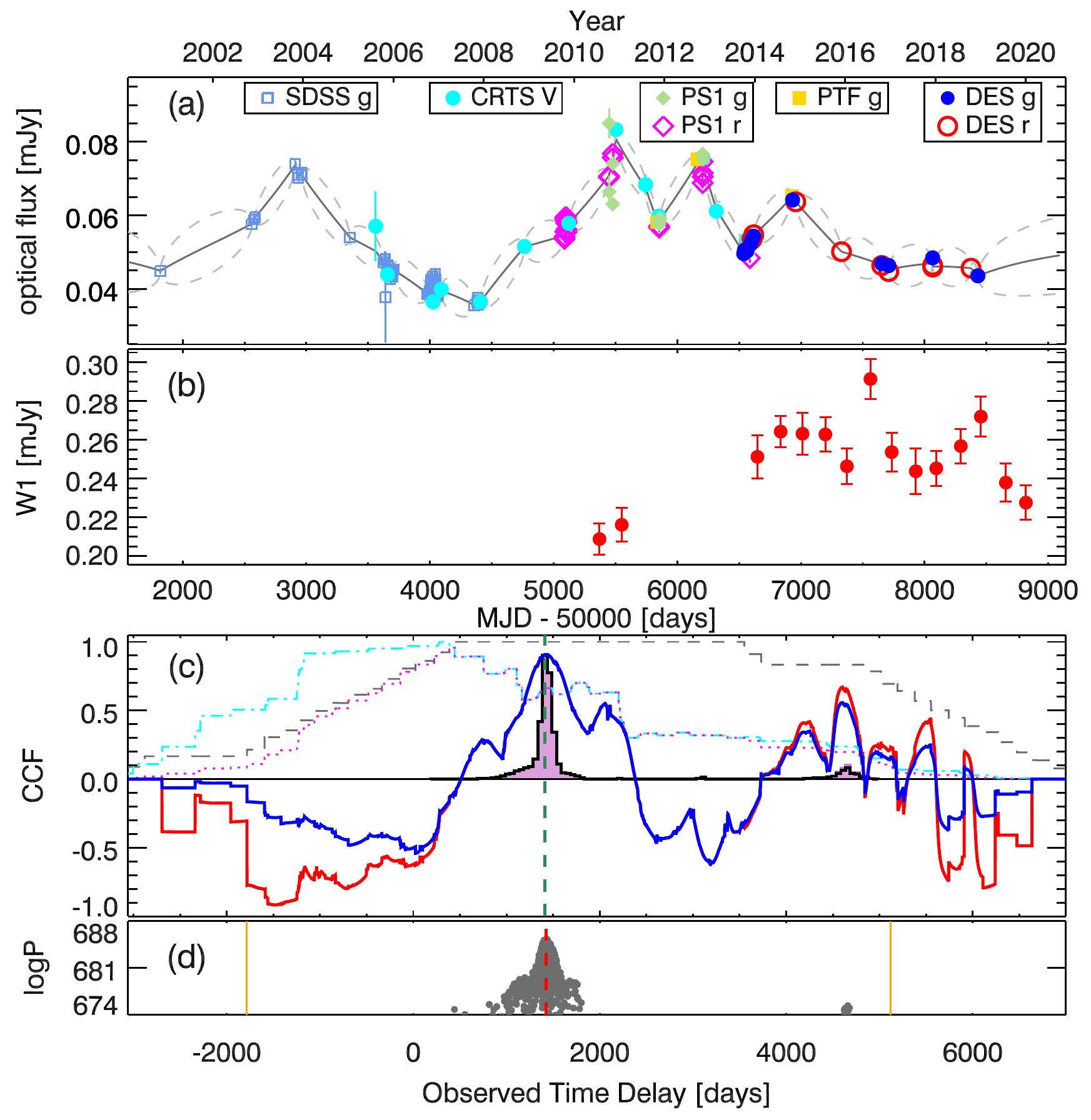}
\includegraphics[width=0.48\textwidth]{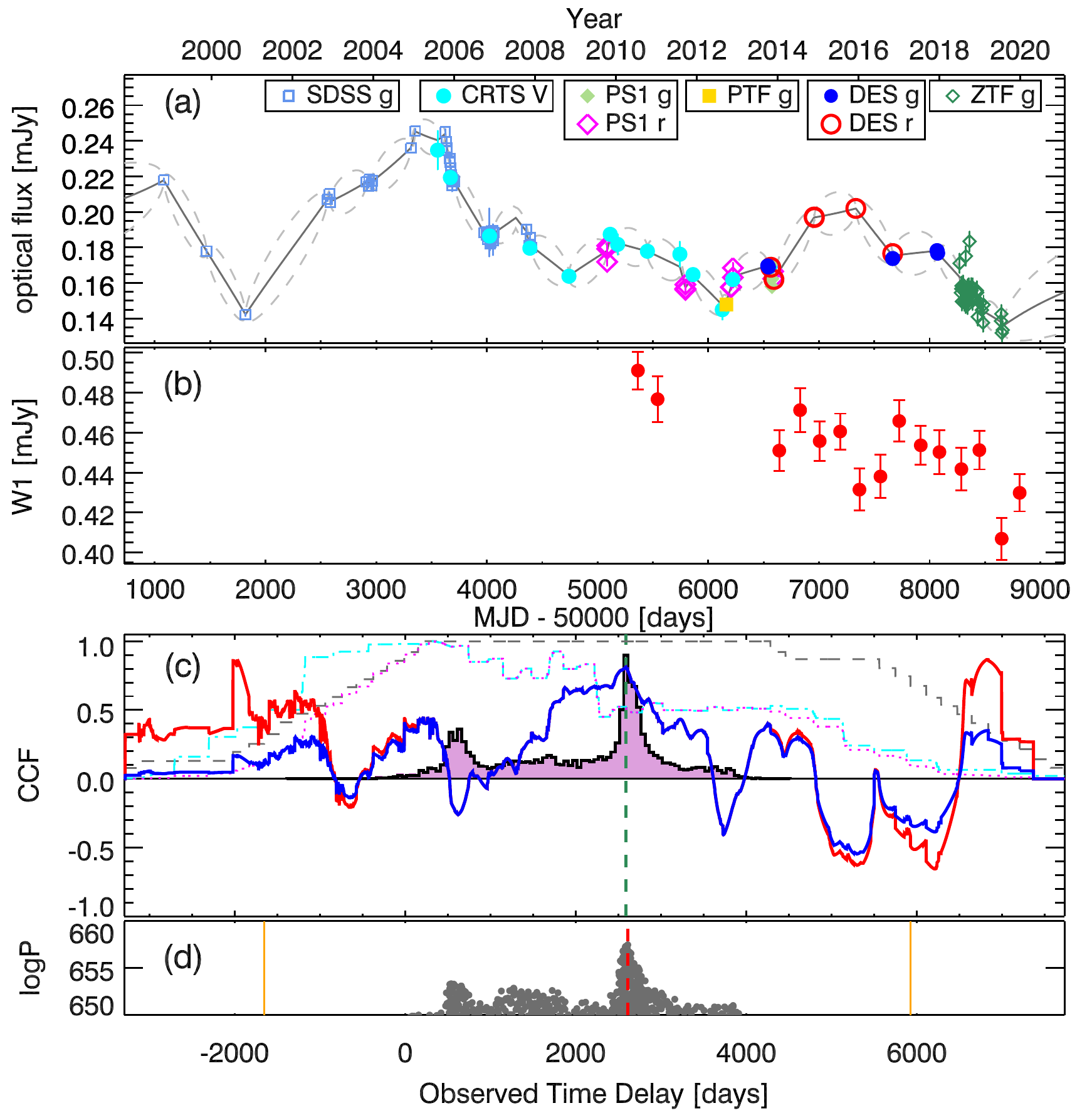}
\caption{Examples of lag detections in our finalcut sample that cover a range of observed-frame lags.
\label{fig:examples}}
\end{figure}

In Figure \ref{fig:examples} we show four more examples of lag detections in our finalcut sample that cover a range of observed-frame lags. The full figure set for all 587 objects in the finalcut sample is available at {http://quasar.astro.illinois.edu/moutai/mir\_lag/}

\bibliography{references.bib,refs.bib}

\end{document}